\pdfoutput=1

\documentclass[
 reprint,
 amsmath,amssymb,
 aps,longbibliography,
 superscriptaddress,
 nofootinbib
]{revtex4-1}

\usepackage{fullpage}
\usepackage[utf8]{inputenc} 
\usepackage{setspace}
\usepackage{graphicx}
\usepackage{dcolumn}
\usepackage{bm}
\usepackage{physics}
\usepackage{multirow}
\usepackage{xcolor}
\usepackage{verbatim}
\usepackage{amsmath}
\usepackage{amssymb}
\usepackage{booktabs}
\usepackage{xcolor}
\usepackage{tikz}
\usepackage{appendix}
\usepackage{makecell}
\usepackage{hyperref}
\usetikzlibrary{quantikz}

\newcommand{\tp}[1]{^{\otimes{#1}}}
\newcommand{\bsm}[1]{$\mathrm{BSM}_{#1}$}
\newcommand{\bsms}[1]{$\mathrm{BSM}_{#1}^{\mathrm{sign}}$}

\newcounter{definition}
\newenvironment{definition}
    {\refstepcounter{definition}\paragraph*{Definition \thedefinition.}}
    {}

\def\P(#1){\Phelper#1|\relax\Pchoice(#1)}
\def\Phelper#1|#2\relax{\ifx\relax#2\relax\def\Pchoice{\Pone}\else\def\Pchoice{\Ptwo}\fi}
\def\Pone(#1){\Pr\qty( #1 )}
\def\Ptwo(#1|#2){\Pr\qty( #1 \,\middle|\, #2 )}
\def\Pr{\mathbf{Pr}}

\begin{document}

\preprint{APS/123-QED}

\title{Loss-tolerant concatenated Bell-state measurement with encoded coherent-state qubits for long-range quantum communication}

\author{Seok-Hyung Lee}
\affiliation{Department of Physics and Astronomy, Seoul National University, Seoul 08826, Republic of Korea}
\author{Seung-Woo Lee}
\affiliation{Center for Quantum Information, Korean Institute of Science and Technology, Seoul 02792, Republic of Korea}
\author{Hyunseok Jeong}
\affiliation{Department of Physics and Astronomy, Seoul National University, Seoul 08826, Republic of Korea}

\date{\today}

\begin{abstract}
    The coherent-state qubit is a promising candidate for optical quantum information processing due to its nearly deterministic nature of the Bell-state measurement (BSM). However, its non-orthogonality incurs difficulties such as failure of the BSM. One may use a large amplitude ($\alpha$) for the coherent state to minimize the failure probability, but the qubit then becomes more vulnerable to dephasing by photon loss. We propose a hardware-efficient concatenated BSM (CBSM) scheme with modified parity encoding using coherent states with reasonably small amplitudes ($|\alpha| \lessapprox 2$), which simultaneously suppresses both failures and dephasing in the BSM procedure. We numerically show that the CBSM scheme achieves a success probability arbitrarily close to unity for appropriate values of $\alpha$ and sufficiently low photon loss rates (e.g., $\lessapprox 5\%$). Furthermore, we verify that the quantum repeater scheme exploiting the CBSM scheme for quantum error correction enables one to carry out efficient long-range quantum communication over 1000 km. We show that the performance is comparable to those of other up-to-date methods or even outperforms them for some cases. Finally, we present methods to prepare logical qubits under modified parity encoding and implement elementary logical operations, which consist of several physical-level ingredients such as generation of superpositions of coherent states (SCSs) and elementary gates under coherent-state basis. Our work demonstrates that the encoded coherent-state qubits in free-propagating fields provide an alternative route to fault-tolerant information processing, especially to long-range quantum communication.
\end{abstract}

\maketitle

\section{Introduction}
\label{sec:introduction}

Optical systems are a competitive candidate for quantum information processing (QIP) due to their long coherence time and advantages in long-distance transmission \cite{kok2010introduction}. It is well known that they are particularly promising for quantum communication. 
Single-photon states are usually considered for the carriers of optical qubits such as vacuum-single-photon-qubit (single-rail encoding) \cite{lund2002nondeterministic} and polarization qubit (dual-rail encoding) \cite{knill2001scheme}. 
However, these encoding schemes have a drawback that the Bell-state measurement (BSM) is non-deterministic with linear optics \cite{lutkenhaus1999bell, calsamiglia2001maximum}. 
The BSM is essential for QIP tasks such as quantum teleportation \cite{bennett1993teleporting, gottesman1999demonstrating} and entanglement swapping \cite{zukowski1993event, pan1998experimental}. 
Quantum teleportation is widely employed not only for quantum communication but also for all-optical quantum computation with gate teleportation \cite{knill2001scheme}. 
It is thus important to overcome the problem of non-deterministic BSM.
Several methods have been suggested using multiple photons for encoding \cite{lee2015nearly, lee2019fundamental}, ancillary states \cite{grice2011arbitrarily, zaidi2013beating, ewert20143}, coherent states \cite{jeong2001quantum, jeong2002purification, jeong2002efficient, ralph2003quantum, glancy2004transmission, braunstein2005quantum, jeong2007schrodinger, lund2008fault, mirrahimi2014dynamically}, and hybrid states \cite{lee2013near, choi2020teleportation, omkar2020resource} to improve the success probability of BSM.
Among them, in this paper, we focus on the scheme using coherent-state qubits that enables one to perform a nearly deterministic BSM with linear optics \cite{jeong2001quantum, jeong2002purification, lee2013bell}. 

Early studies on coherent states as carriers of quantum information focus on how to construct logical qubits and elementary logical gates  \cite{cochrane1999macroscopically, lloyd1999quantum, de2000quantum, jeong2001quantum, jeong2002efficient, bartlett2002quantum, ralph2003quantum}. 
In these works, the basis set is chosen either as $\qty{\ket{\pm\alpha}}$ or as $\qty{ N_\pm \qty( \ket{\alpha} \pm \ket{-\alpha} )}$, where $\ket{\pm\alpha}$ are coherent states of amplitudes $\pm\alpha$ and $N_\pm$ are normalization factors.
Various attempts to obtain fault-tolerance on QIP with coherent states have been made, starting from simple embedding on well-known discrete-variable (DV) encoding schemes \cite{glancy2004transmission, lund2008fault}, to exploiting the property of continuous-variable (CV) systems \cite{leghtas2013hardware, mirrahimi2014dynamically, albert2016holonomic, puri2017engineering, li2017cat, cohen2017degeneracy, puri2019stabilized}, with some experimental demonstration \cite{leghtas2015confining, ofek2016extending, touzard2018coherent, rosenblum2018fault}. 
Recently, it was claimed that simple 1D repetition cat code enables hardware-efficient topologically-protected quantum computation by exploiting the 2D phase space for logical operations \cite{guillaud2019repetition}. 
However, these studies mainly deal with coherent-state qubits inside a cavity system, and they cannot be directly applied to fault-tolerant QIP in free-propagating optical fields. 
Our main goal is to investigate the possibility to use simple concatenated repetition codes, which can be generated and manipulated with combinations of well-known elementary gates, for fault-tolerant QIP with free-propagating coherent-state qubits.

As mentioned before, the BSM of coherent-state qubits, where the basis is $\qty{ \ket{\pm\alpha} }$, is nearly deterministic. 
However, due to the non-orthogonality of the basis set, a small but non-negligible probability of failure exists \cite{jeong2001quantum, jeong2002purification}. 
One may use coherent states with large values of $\alpha$ to solve this problem, but the qubit then becomes more vulnerable to dephasing by photon loss \cite{glancy2004transmission}. 
It is impossible to ideally suppress both failures and dephasing simultaneously with such an elementary coherent-state encoding. 
In this paper, motivating by recent works on \textit{concatenated Bell-state measurement} (CBSM) with multi-photon polarization qubits \cite{muralidharan2014ultrafast, lee2019fundamental} and repetition cat code \cite{guillaud2019repetition}, we overcome these obstacles by introducing the CBSM with \textit{modified parity encoding} employing coherent states. 
We propose an elaborately designed CBSM scheme with consideration of hardware-efficiency, and numerically show that the scheme successfully suppresses both failures and dephasing simultaneously with reasonably small amplitudes (e.g., $|\alpha| < 2$) of coherent states.

One of the key applications with BSMs is long-distance quantum communication through quantum repeaters \cite{sangouard2011quantum}.
In the initially proposed quantum repeater schemes to generate Bell pairs between distant parties \cite{briegel1998quantum, dur1999quantum, duan2001long, kok2003construction, simon2007quantum}, heralded entanglement generation is required for suppressing transmission errors, which makes long-lived quantum memory essential \cite{muralidharan2016optimal}. 
Recently, quantum repeater schemes exploiting quantum error correction (QEC) have been suggested for suppressing errors,  which do not require long-lived quantum memory, have been suggested \cite{jiang2009quantum, munro2010quantum, sangouard2010quantum, munro2012quantum, muralidharan2014ultrafast, azuma2015all, muralidharan2016optimal, zwerger2016measurement, ewert2016ultrafast, lee2019fundamental}, where a quantum repeater is built up without long-lived quantum memory by encoding information with QEC codes, sending it by lossy channel, and relaying the encoded information from each station to the next station with error corrections. 
In each repeater station, a fault-tolerant BSM can be used for QEC by teleporting the incoming lossy logical qubits with a generated logical Bell state \cite{lee2019fundamental}. 
Later in this paper, we evaluate the performance of the quantum repeater scheme using our CBSM scheme and show that it indeed enables quantum repeater with high performance over distances longer than 1000 km.

The outline of the paper is as follows. 
In Sec.~\ref{sec:bsm_coherent_state_qubits}, we review the BSM scheme of lossless coherent-state qubits, extend it to lossy cases, and evaluate its success, failure, and error rates against the coherent-state amplitude $\alpha$ and the photon survival rate. 
In Sec.~\ref{sec:parity_encoding_scheme}, we present the modified parity encoding scheme employing coherent-state qubits, and show the hierarchy relation between logical, block, and physical level. 
In Sec.~\ref{sec:cbsm_with_coherent_state_qubits}, we first suggest an unoptimized CBSM scheme which only uses simple majority votes and counting of measurement results, and analyze the root of fault-tolerance of the scheme. 
After that, we propose an improved CBSM scheme which is elaborately designed considering hardware efficiency. 
In Sec.~\ref{sec:prob_dist_of_measurement_results}, we present the analytic expressions of the probability distributions of CBSM results, which are simple matrix forms enabling fast sampling of the results and can be generalized to any CBSM schemes. 
In Sec.~\ref{sec:numerical_calculations}, we show the results of numerical calculations. We first present a performance analysis by the success, failure, and error probabilities of CBSM. 
We then investigate the performance of the quantum repeater scheme which uses our CBSM scheme for error correction, as one of the key applications of BSM. 
In Sec.~\ref{sec:implementation}, we describe methods to prepare the logical qubits under modified parity encoding and implement elementary logical operations, which consist of several physical-level ingredients such as generation of superpositions of coherent states (SCSs) and elementary gates under coherent-state basis. 
We also briefly review recent progresses on realizations of these ingredients. 
We conclude with final remarks in Sec.~\ref{sec:conclusion}

\section{Bell-state measurement of lossy coherent-state qubits}
\label{sec:bsm_coherent_state_qubits}

\begin{figure}[tb]
	\centering
	\includegraphics[width=0.6\linewidth]{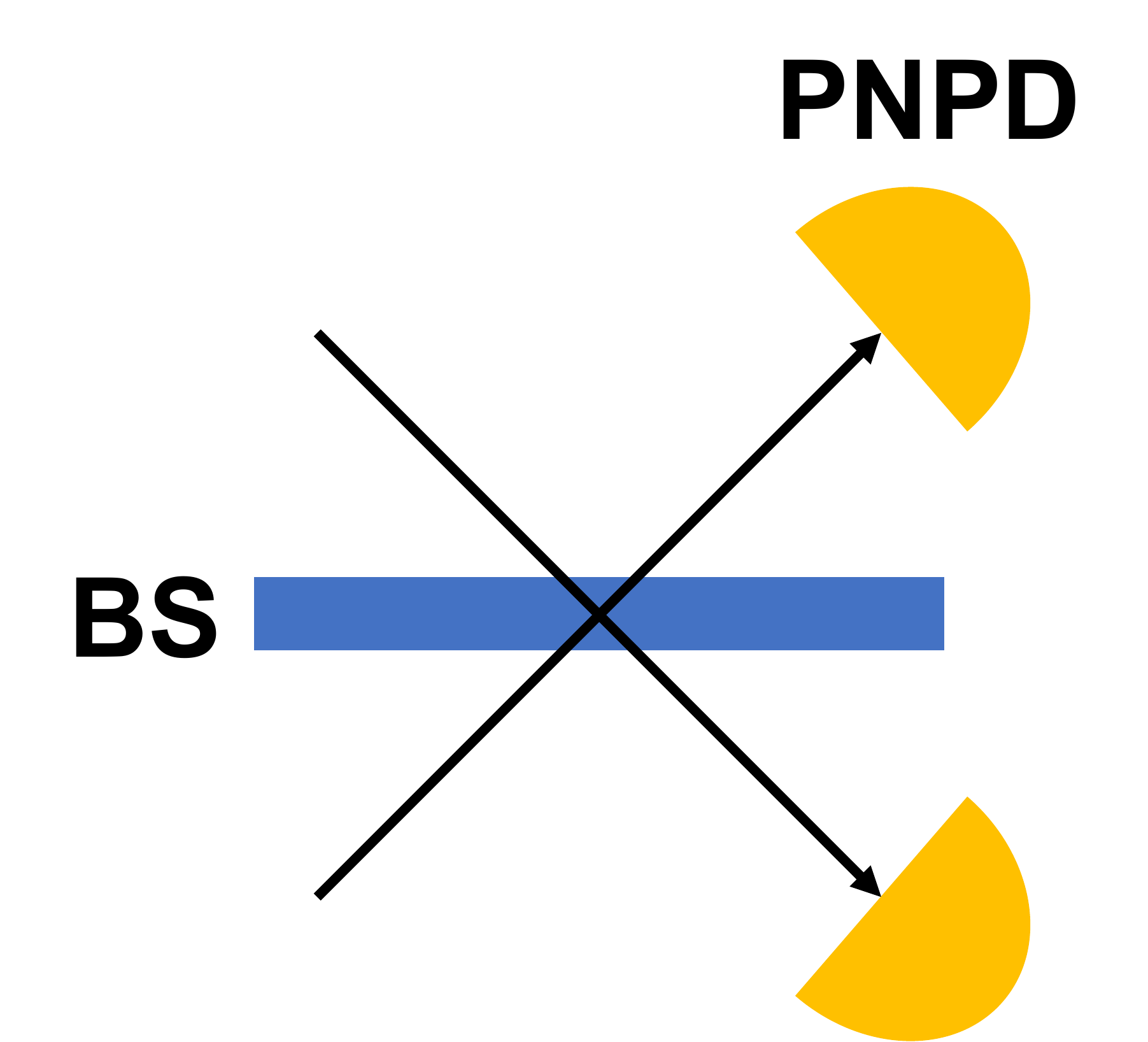}
	\caption{The BSM scheme of coherent-state qubits \cite{jeong2001quantum}. It uses one 50:50 beam splitter (BS) and two photon-number parity detectors (PNPDs). The result is determined by the measurement results of the PNPDs as Eq.~\eqref{eq:interpreting_pnpd_physical_bsm}.}
	\label{fig:bsm_coherent_state_qubits}
\end{figure}

We first review the BSM scheme of lossless coherent-state qubits encoded with basis
\begin{align}
    \ket{0_L} := \ket{\alpha}, ~~ \ket{1_L} := \ket{-\alpha}.
    \label{eq:coherent_state_qubit}
\end{align}
The four Bell states of coherent-state qubits are
\begin{align*}
    \ket{\phi_\pm} &:= N_\pm \qty( \ket{\alpha}\ket{\alpha} \pm \ket{-\alpha}\ket{-\alpha} ), \\
    \ket{\psi_\pm} &:= N_\pm \qty( \ket{\alpha}\ket{-\alpha} \pm \ket{-\alpha}\ket{\alpha} ),
\end{align*}
where $N_\pm := \qty[ 2\qty( 1 \pm e^{-4|\alpha|^2} ) ]^{-1/2}$ are normalization factors.
A BSM of lossless coherent-state qubits is performed with a 50:50 beam splitter and two photon number parity detectors (PNPDs) \cite{jeong2001quantum, jeong2002purification}, as seen in Fig.~\ref{fig:bsm_coherent_state_qubits}.
The four Bell states can be deterministically identified from the results of the PNPDs unless both of the PNPDs do not detect any photons as 
\begin{align}
    (\mathrm{even}, 0) \rightarrow \ket{\phi_+},~~ & (\mathrm{odd}, 0) \rightarrow \ket{\phi_-}, \nonumber \\
    (0, \mathrm{even}) \rightarrow \ket{\psi_+},~~ & (0, \mathrm{odd}) \rightarrow \ket{\psi_-}.
    \label{eq:interpreting_pnpd_physical_bsm}
\end{align}
In the case that both of the PNPDs do not detect photons, which we call `failure,' only the sign ($\pm$ for $\ket{\phi_\pm}$ and $\ket{\psi_\pm}$) of the Bell state can be determined since there exists ambiguity between $\ket{\phi_+}$ and $\ket{\psi_+}$.

For realistic scenarios, we need to introduce photon loss.
We use the photon loss model by the Master equation under the Born-Markov approximation with zero temperature \cite{phoenix1990wave}:
\begin{equation}
    \frac{\partial \rho}{\partial \tau} =  \gamma \sum_i \qty( \hat{a}_i \rho \hat{a}_i^\dagger -\frac{1}{2}\hat{a}_i^\dagger \hat{a}_i \rho -\frac{1}{2}\rho \hat{a}_i^\dagger \hat{a}_i ), \label{eq:photon_loss_master_equation}
\end{equation}
where $\rho(\tau)$ is the density operator of system suffering photon loss as the function of time $\tau$, $\gamma$ is the decay constant, and $\hat{a}_i$ ($\hat{a}_i^\dagger$) is the annihilation (creation) operator of the $i$th mode. 
It is known that this photon loss model is equivalent with the beam splitter model where each mode is independently mixed with the vacuum state by a beam splitter with the transmittance $t = e^{-\gamma\tau/2}$ and the reflectance $r = \sqrt{1 - t^2}$ \cite{leonhardt1993quantum}:
\begin{equation}
    \begin{pmatrix}
        \hat{a} \\ \hat{b}
    \end{pmatrix} 
    \rightarrow
    \begin{pmatrix}
        \hat{a}' \\ \hat{b}'
    \end{pmatrix}
    = 
    \begin{pmatrix}
        t & -r \\
        r & t
    \end{pmatrix}
    \begin{pmatrix}
        \hat{a} \\ \hat{b}
    \end{pmatrix}.\label{eq:photon_loss_beam_splitter}
\end{equation}
Here, $\hat{a}$ ($\hat{a}'$) is the annihilation operator of the input (output) mode, and $\hat{b}$ ($\hat{b}'$) is that of the input (output) mode of the ancillary system which is initially in the vacuum state.
The final state after suffering photon loss is obtained by tracing out the ancillary system from the output state of the beam splitter.
Considering the photon survival rate $\eta = t^2$, the final state can be expressed in terms of $\eta$.

Now, we consider the BSM on lossy coherent-state qubits.
Precisely speaking, we deal with a situation that the two coherent-state qubits suffer photon losses before the BSM of Fig.~\ref{fig:bsm_coherent_state_qubits} is performed.
We first rewrite each element of the BSM scheme in mathematical term: $\mathcal{U}_{\mathrm{BS}}$ is a unitary channel corresponding to a 50:50 beam splitter, $\mathit{\Lambda}_{\eta}$ is a photon loss channel with a survival rate $\eta$, and $\Pi_x$ for $x \in \qty{ 0, 1, 2 }$ is a projector defined by
\begin{align*}
    &\Pi_0 := \dyad{0_\mathrm{F}}, \quad \Pi_1 := \sum_{n \geq 1:\mathrm{odd}} \dyad{n_\mathrm{F}},\\
    &\Pi_2 := \sum_{n \geq 2:\mathrm{even}} \dyad{n_\mathrm{F}},
\end{align*}
where $\ket{n_\mathrm{F}}$ is the Fock state with a photon number of $n$.
A set of operators,
\begin{equation*}
    M_{x,y} := \qty[ \mathcal{U}_{\mathrm{BS}} \circ \qty( \mathit{\Lambda}_{\eta_1} \otimes \mathit{\Lambda}_{\eta_2} ) ]^\dagger \qty( \Pi_x \otimes \Pi_y )
\end{equation*}
with $x, y \in \{ 0, 1, 2 \}$, then forms a positive-operator valued measure (POVM) corresponding to the BSM of lossy coherent-state qubits.
Explicit forms of them are presented in Appendix \ref{app:povm_elements}.

\begin{table}[b]
    \caption{
        Correspondences between the pairs of the PNPD results and the resulting Bell states. 
        The Bell state $\ket{B} \in \qty{ \ket{\phi_\pm}, \ket{\psi_\pm} }$ is chosen to maximize the posterior probability $\mathbf{Pr}\qty( B \middle| x, y )$. 
        Here, $x$ and $y$ indicate the results of two PNPDs, where 0, 1, and 2 mean zero, odd, and even detection, respectively.
        The cases that both $x$ and $y$ are nonzero occur only when the loss rates of the two modes are different.
        We also note that only the sign of the Bell state can be determined in the cases of $x=y$, which we call 'failure,' since both $\ket{\phi_+}$ and $\ket{\psi_+}$ maximize the posterior probability at the same time.
    }
    \label{table:interpreting_single_BSM_results}
    \centering
    \begin{ruledtabular}
        \begin{tabular}{cccc}
            $x$ \textbackslash{} $y$ & 0                 & 1                 & 2                 \\ \hline
            0                        & $\phi_+$ or $\psi_+$
            & $\psi_-$          & $\psi_+$          \\ 
            1                        & $\phi_-$          & $\phi_+$ or $\psi_+$ & $\psi_-$          \\ 
            2                        & $\phi_+$          & $\phi_-$          & $\phi_+$ or $\psi_+$
        \end{tabular}
    \end{ruledtabular}
\end{table}

Assuming the equal prior probability distribution of the four Bell states $\mathcal{B}_0 = \qty{ \ket{\phi_\pm}, \ket{\psi_\pm} }$, we choose the Bell state $\ket{B} \in \mathcal{B}_0$ which maximizes the posterior probability from the PNPD results $(x, y)$:
\begin{align}
    \P(B|x, y) &= \frac{\P(x, y|B) \P(B)}{\sum_{\ket{B'} \in \mathcal{B}_0} \P(x, y|B') \P(B')} \nonumber \\
    &\propto \P(x, y|B) = \bra{B} M_{x,y} \ket{B}.
    \label{eq:single_bsm_cond_prob}
\end{align}
In other words, we choose $\ket{B} \in \mathcal{B}_0$ satisfying
\begin{align}
    \ket{B} = \underset{\ket{B'} \in \mathcal{B}_0}{\mathrm{argmax}} \bra{B'} M_{x,y} \ket{B'}, \label{eq:single_bsm_choosing_state}
\end{align}
for the final result of the BSM.
A straightforward analysis with Eq.~\eqref{eq:single_bsm_choosing_state} and the POVM elements of BSM presented in Appendix \ref{app:povm_elements} shows the correspondences between the pairs of the PNPD results and the resulting Bell states as shown in Table \ref{table:interpreting_single_BSM_results}.
We note that, when losses are considered, there are some cases that never happen for lossless cases. 
In other words, both $x$ and $y$ can be nonzero at the same time, while the probabilities of these cases vanish for $\eta_1 = \eta_2$.

If the state before suffering the photon loss is one of the four Bell states, there are five possible cases regarding the result of the measurement: success, $X$-error, $Z$-error, $Y$-error, and failure.
If the resulting Bell state is the same with the initial one, we call it success.
$X$-error corresponds to `letter flip', i.e., the change of the letter ($\phi$ or $\psi$) in a Bell state such as from $\ket{\phi_+}$ to $\ket{\psi_+}$. 
$Z$-error corresponds to `sign flip', i.e., the change of the sign ($\pm$) in a Bell state such as from $\ket{\phi_+}$ to $\ket{\phi_-}$.
$Y$-error corresponds to simultaneous symbol and sign flips.
The last case, failure, corresponds to the cases of $x=y$ in Table~\ref{table:interpreting_single_BSM_results} that the letter of the Bell state cannot be determined since both $\ket{\phi_+}$ and $\ket{\psi_+}$ maximize the posterior probability at the same time.
We would like to emphasize that the sign still can be determined for this case.

\begin{figure}[tb]
    \centering
    \includegraphics[width=\linewidth]{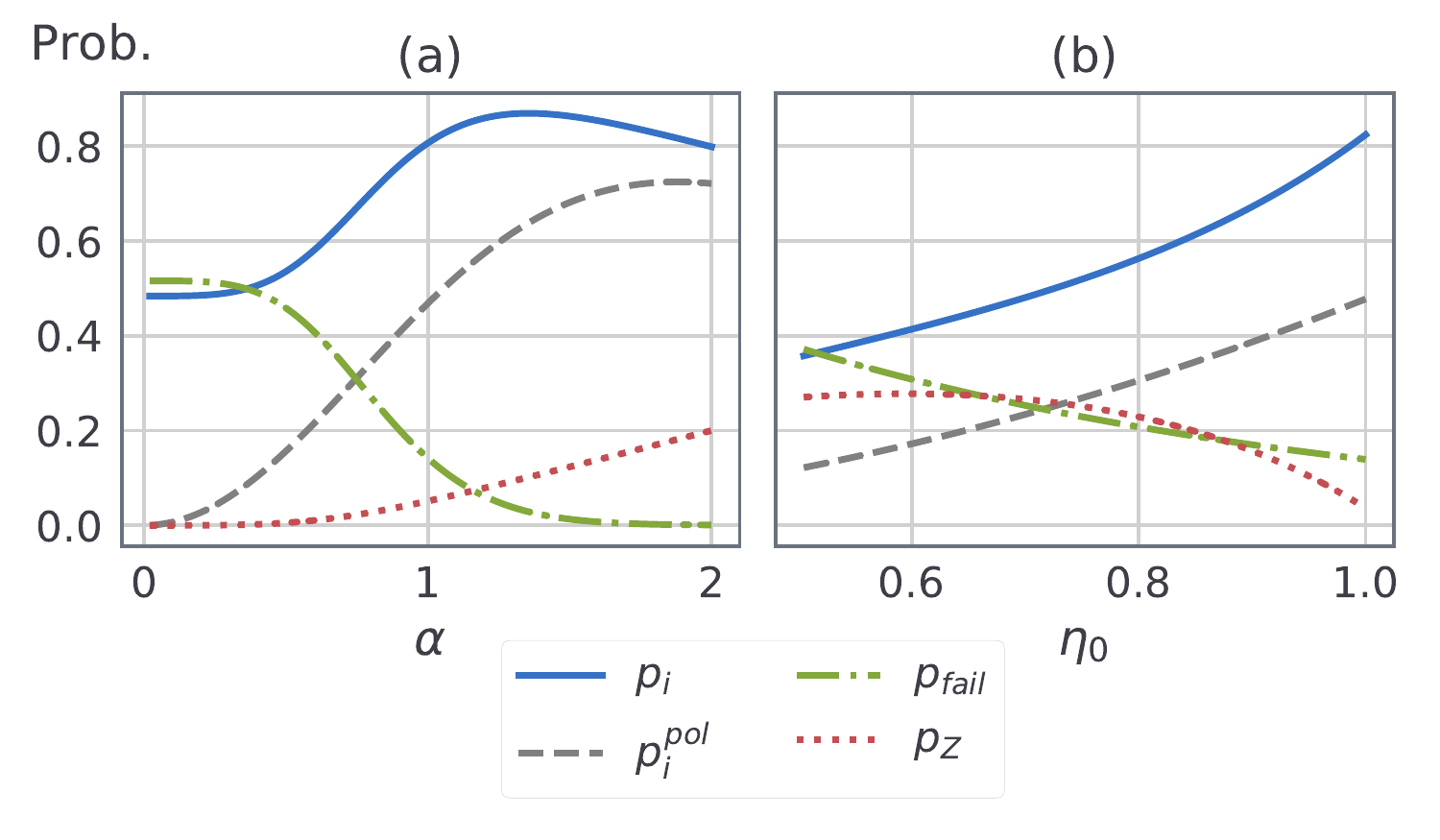}
    \caption{The success ($p_i$), failure ($p_{fail}$), and $Z$-error probabilities ($p_z$) of BSM on coherent-state qubits against (a) $\alpha$ (fixing $\eta_0=0.99$) (b) $\eta_0$ (fixing $\alpha = 1$). We set the photon survival rates of the two systems as $\eta_1 := \eta_0$ and $\eta_2 := \eta_0 e^{-L/L_{\mathrm{att}}}$, where $L := 1~\text{km}$ and $L_{\mathrm{att}} := 22 \text{ km}$. It corresponds to the situation that both systems suffer internal losses with the photon survival rates of $\eta_0$ and the photons of the second system travel the distance of $L := 1~\text{km}$ before the measurement. The blue solid line is the success probability $p_i$, the green dash-dotted line is the failure probability $p_{fail}$, and the red dotted line is the $Z$-error probability $p_z$. Also, the gray dashed line is the success probability of BSM on multi-photon polarization qubits for different photon numbers \cite{lee2015nearly}, which is plotted for comparison, where $\alpha$ is now the amplitude of the coherent state which has the same photon number with the qubit. The $X$-error ($p_x$) and $Y$-error probabilities ($p_y$) are not plotted since they are much smaller than other probabilities regardless of $\alpha$ and $\eta_0$: $p_x, p_y \lessapprox 10^{-4}$}
    \label{fig:phys_bsm_probs}
\end{figure}

Now, we numerically analyze the success, failure, and error probabilities of BSM on coherent-state qubits. We consider a BSM on coherent-state qubits performed jointly on two systems which suffer internal losses with the survival rates of $\eta_0$ and the photons of the second system travel the distance of $L = 1~\text{km}$ before the measurement.
The photon survival rates of the two systems are then $\eta_1 := \eta_0$ and $\eta_2 := \eta_0 e^{-L_0/L_{\mathrm{att}}}$, respectively, where $L_{\mathrm{att}} = 22~\text{km}$ is the attenuation length.

Figure~\ref{fig:phys_bsm_probs} shows the success, failure, and error probabilities of the BSM in this situation against the amplitude $\alpha$ of the coherent state and the internal photon survival rate $\eta_0$. 
It shows the well-known fact that the success probability is higher than that of a BSM on multi-photon polarization qubits with the same photon number. 
Also, the failure and $Z$-error probabilities have a trade-off relation with changing $\alpha$; when $\alpha$ increases, failures get less probable while $Z$-errors get more probable. 
It is because coherent states with large amplitudes have less overlaps with the vacuum state and are more vulnerable to dephasing by photon loss. 
Furthermore, we would like to emphasize that the error is strongly biased, i.e., the $X$- and $Y$-error probabilities are much smaller than the failure and $Z$-error probabilities regardless of $\alpha$ and $\eta_0$: $p_x, p_y \lessapprox 10^{-4}$. 
They even vanish if $\eta_1 = \eta_2$, which is the consequence from the fact that both $x$ and $y$ in Table~\ref{table:interpreting_single_BSM_results} can be nonzero simultaneously only when the two photon survival rates are different.
This fact is important for constructing a hardware-efficient CBSM scheme in Sec.~\ref{subsec:cbsm_scheme_with_optimized_cost}.

\section{Modified parity encoding scheme with coherent-state qubits}
\label{sec:parity_encoding_scheme}

Now, we present the encoding scheme we use for our CBSM scheme. We modify the parity state encoding or generalized Shor's encoding \cite{ralph2005loss, lee2019fundamental} for the coherent-state qubit. The modified parity encoding is defined as follows.

\begin{definition}
\label{def:parity_encoding}
The basis qubits $\qty{ \ket{0_L}, \ket{1_L} }$ of $(n, m, \alpha)$ \textit{modified parity encoding} scheme where $n$ and $m$ are odd integers and $\alpha$ is a complex number are defined as:
\begin{align*}
    \ket{0_L} &:= \qty[ N^{(m)} \qty{ \ket{\widetilde{+}}\tp{m} + \ket{\widetilde{-}}\tp{m} } ]\tp{n}, \\
    \ket{1_L} &:= \qty[ N^{(m)} \qty{ \ket{\widetilde{+}}\tp{m} - \ket{\widetilde{-}}\tp{m} } ]\tp{n},
\end{align*}
where $\ket{\widetilde{\pm}} := \ket{\alpha} \pm \ket{-\alpha}$ are unnormalized SCSs (we use the tilde above the ket to denote that it is unnormalized) and $N^{(m)} := \qty[ 2^m \qty{ \qty( 1 + e^{-2|\alpha|^2} )^m + \qty( 1 - e^{-2|\alpha|^2} )^m } ]^{-1/2}$. This encoding scheme coincides the original coherent-state encoding in Eq.~\eqref{eq:coherent_state_qubit} when $n = m = 1$.
\end{definition}

The modified parity encoding has a hierarchy structure of Hilbert spaces: \textit{logical}, \textit{block}, and \textit{physical} level. The \textit{logical-level space} is the total Hilbert space spanned by $\qty{ \ket{0_L}, \ket{1_L} }$. It can be divided into $n$ \textit{block-level spaces} (referred as blocks), each of which is spanned by $\qty{ \ket{\pm^{(m)}} }$ where $\ket{\pm^{(m)}} := N^{(m)} \qty{ \qty( \ket{\tilde{+}} )\tp{m} \pm \qty( \ket{\tilde{-}} )\tp{m} }$. A block is again divided into $m$ \textit{physical-level spaces} (referred as PLSs), each of which is spanned by $\ket{\pm\alpha}$.

We also define four Bell states for each level as following, where normalization constants are omitted:

\paragraph{Logical level:}
\begin{align*}
    \ket{\Phi_\pm} &:= \ket{0_L}\ket{0_L} \pm \ket{1_L}\ket{1_L} \\
    \ket{\Psi_\pm} &:= \ket{0_L}\ket{1_L} \pm \ket{1_L}\ket{0_L}
\end{align*}

\paragraph{Block level:}
\begin{align*}
    \ket{\phi^{(m)}_\pm} &:= \ket{+^{(m)}}\ket{+^{(m)}} \pm \ket{-^{(m)}}\ket{-^{(m)}} \\
    \ket{\psi^{(m)}_\pm} &:= \ket{+^{(m)}}\ket{-^{(m)}} \pm \ket{-^{(m)}}\ket{+^{(m)}}
\end{align*}

\paragraph{Physical level:}
\begin{align*}
    \ket{\phi_\pm} &:= \ket{\alpha}\ket{\alpha} \pm \ket{-\alpha}\ket{-\alpha} \\
    \ket{\psi_\pm} &:= \ket{\alpha}\ket{-\alpha} \pm \ket{-\alpha}\ket{\alpha}
\end{align*}

Each logical-level Bell state can be decomposed into block-level Bell states:
\begin{subequations}
\label{eqs:logical_to_block}
\begin{align}
    &\ket{\Phi_{+(-)}} = \tilde{N}_{\pm, n, m} \nonumber \\
    &\quad \times \sum_{k=\text{even(odd)} \leq n} \mathcal{P}\qty[ \ket{\widetilde{\phi^{(m)}_-}}^{\otimes k} \ket{\widetilde{\phi^{(m)}_+}}^{\otimes n-k} ], \label{eq:logical_to_block_phi} \\
    &\ket{\Psi_{+(-)}} = \tilde{N}_{\pm, n, m} \nonumber \\
    &\quad \times \sum_{k=\text{even(odd)} \leq n} \mathcal{P}\qty[ \ket{\widetilde{\psi^{(m)}_-}}^{\otimes k} \ket{\widetilde{\psi^{(m)}_+}}^{\otimes n-k} ],
\end{align}
\end{subequations}
where
\begin{align}
    \tilde{N}_{\pm, n, m} &:= \frac{1}{\sqrt{2^{n-1}}} \qty[ 1 \pm u(\alpha, m)^{2n} ]^{-\frac{1}{2}}, \label{eq:N_tilde_def}\\
    \ket{\widetilde{\phi^{(m)}_\pm}} &:= \qty[ 1 \pm u(\alpha, m)^2 ]^\frac{1}{2} \ket{\phi^{(m)}_\pm}, \label{eq:phi_tilde_def} \\
    \ket{\widetilde{\psi^{(m)}_\pm}} &:= \qty[ 1 \pm u(\alpha, m)^2 ]^\frac{1}{2} \ket{\psi^{(m)}_\pm}, \nonumber \\
    u(\alpha, m) &:= \frac{ \qty( 1 + e^{-2|\alpha|^2} )^m - \qty( 1 - e^{-2|\alpha|^2} )^m }{ \qty( 1 + e^{-2|\alpha|^2} )^m + \qty( 1 - e^{-2|\alpha|^2} )^m } \label{eq:u_def},
\end{align}
and $\mathcal{P}[\cdot]$ is the summation of all the possible permutations of the tensor product inside the square bracket.

Similarly, each block-level Bell state can be decomposed into physical-level Bell states:
\begin{subequations}
\label{eqs:block_to_physical}
\begin{align}
    \ket{\phi^{(m)}_{\pm}} &= \frac{\tilde{N}_{\pm, 1, m}}{\sqrt{2}} \sum_{l=\text{even} \leq m} \mathcal{P}\qty[ \ket{\psi_\pm}^{\otimes l} \ket{\phi_\pm}^{\otimes m-l} ], \label{eq:block_to_physical_phi} \\
    \ket{\psi^{(m)}_{\pm}} &= \frac{\tilde{N}_{\pm, 1, m}}{\sqrt{2}} \sum_{l=\text{odd} \leq m} \mathcal{P}\qty[ \ket{\psi_\pm}^{\otimes l} \ket{\phi_\pm}^{\otimes m-l} ]. \label{eq:block_to_physical_psi}
\end{align}
\end{subequations}

The core of CBSM is contained in Eqs.~\eqref{eqs:logical_to_block} and \eqref{eqs:block_to_physical}; they make it possible to perform a logical BSM by the combination of $n$ block-level BSMs, each of which is again performed by the combination of $m$ physical-level BSMs.

The equations also show that, in a lossless system, a CBSM does not incur any logical error, i.e., the only possible cases are success and failure. This property is important since failures are detectable whereas logical errors are not. Hence, the modified parity encoding in \textit{Definition \ref{def:parity_encoding}} is the natural extension of the original coherent-state encoding in Eq.~\eqref{eq:coherent_state_qubit}, in the sense that this desired property still remains. If we use other states such as normalized SCSs or coherent states in place of unnormalized SCSs $\ket{\widetilde{\pm}}$ for the encoding, this property no longer exists.

\section{Concatenated Bell-state measurement with encoded coherent-state qubits}
\label{sec:cbsm_with_coherent_state_qubits}

\begin{figure}
    \centering
    \includegraphics[width=\columnwidth]{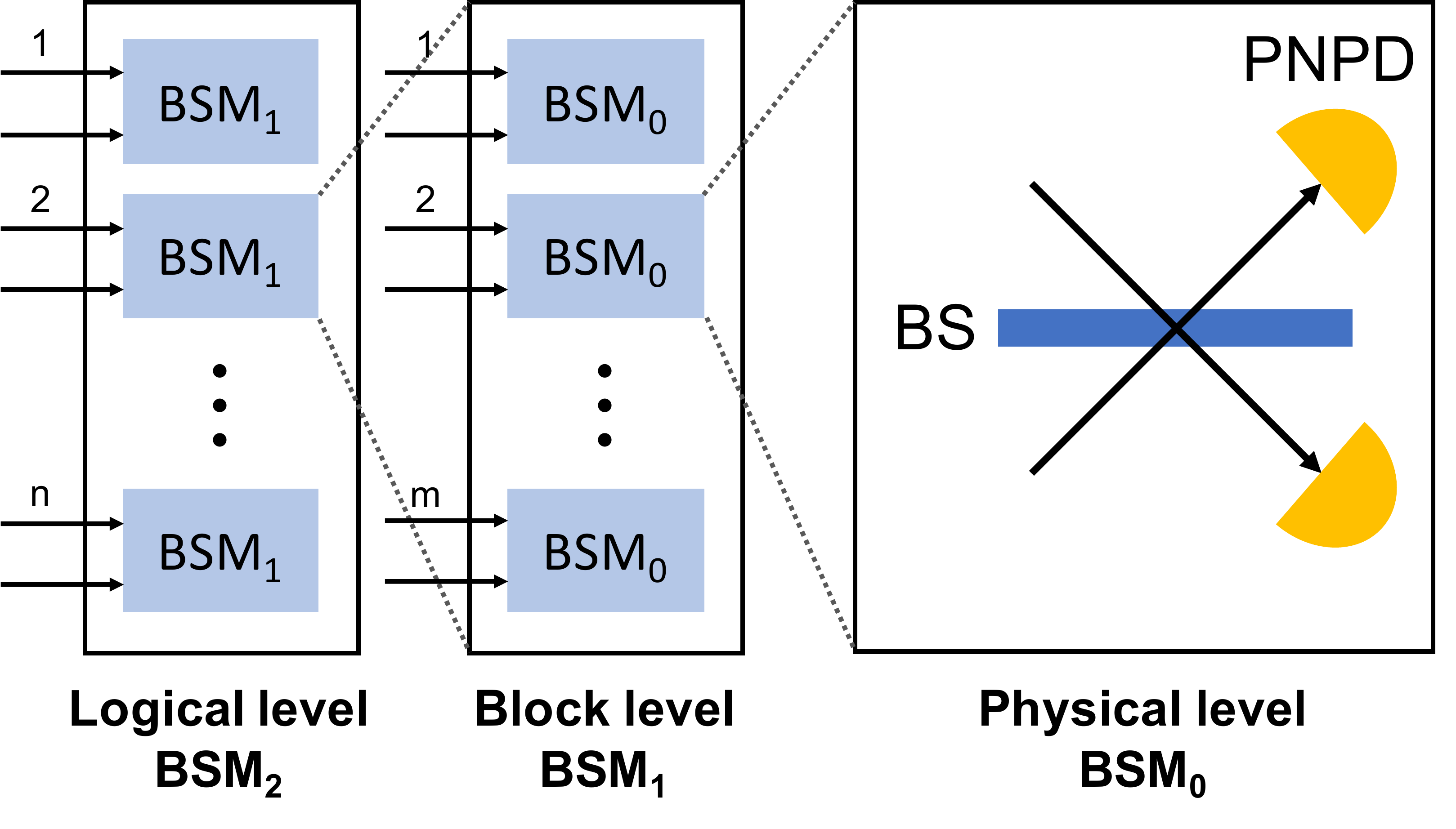}
    \caption{Schematic figure of CBSM schemes with coherent-state qubits. The scheme is done in concatenated manner: each logical-level BSM (\bsm{2}) is done by the combination of $n$ block-level BSMs (\bsm{1}). Each block-level BSM is again done by the combination of $m$ physical-level BSMs (\bsm{0}).}
    \label{fig:cbsm_schematic}
\end{figure}

Now, we suggest \textit{concatenated Bell-state measurement} (CBSM) schemes with the modified parity encoding presented in the previous section. The schematic figure of the CBSM schemes is shown in Fig.~\ref{fig:cbsm_schematic}. As mentioned in the previous section, each logical-level BSM is done by the composition of $n$ block-level BSMs and each block-level BSM is done by the composition of $m$ physical-level BSMs. We first consider an unoptimized scheme which consists of simple counting of measurement results. We then present a hardware-efficient scheme which can significantly reduce the expected cost of the CBSM defined in terms of the expected number of physical-level BSMs used for a single CBSM.

\subsection{Unoptmized CBSM scheme}
\label{subsec:unoptimized_cbsm_scheme}

\begin{table*}[tb]
    \caption{Interpretation of the measurement results in the unoptimized CBSM scheme. It is also valid in the hardware-efficient CBSM scheme, if we consider the results of \bsm{0} (\bsm{1}) and \bsms{0} (\bsms{1}) together when determining the sign of the block (logical) level Bell state.}
    \label{table:unoptimized_cbsm_scheme}
    \begin{ruledtabular}
    \begin{tabular}{ccc}
        Level & Sign ($\pm$) & Letter ($\phi$ or $\psi$) \\ \hline
        \makecell{Physical\\(\bsm{0})} & \multicolumn{2}{c}{BSM scheme of original coherent-state qubits} \\
        \makecell{Block\\(\bsm{1})} & \makecell{Majority vote of the signs \\ of the \bsm{0} results} & \makecell{Number of \bsm{0} results with $(-)$ sign: \\ $\phi$ if even, $\psi$ if odd} \\
        \makecell{Logical\\(\bsm{2})} & \makecell{Number of \bsm{1} results with $\psi$ letter: \\ $(+)$ if even, $(-)$ if odd} & \makecell{Majority vote of the letters \\ of the \bsm{1} results}
    \end{tabular}
    \end{ruledtabular}
\end{table*}

Here, we suggest a CBSM scheme which is unoptimized but much simpler than the hardware-efficient scheme presented in the next subsection. It is straightforward to justify the scheme with Eqs.~\eqref{eqs:logical_to_block} and \eqref{eqs:block_to_physical}. The interpretation of the measurement results in the scheme is summarized in Table \ref{table:unoptimized_cbsm_scheme}.

\subsubsection{Physical level: \bsm{0}}

For a physical-level BSM (referred as \bsm{0}), we use the BSM scheme for single lossy coherent-state qubit presented in Fig.~\ref{fig:bsm_coherent_state_qubits} and Table \ref{table:interpreting_single_BSM_results}. Remark that the sign of the Bell state is always determinable, while its letter is not determinable if the results of the two PNPDs are the same, i.e., $x = y$ in Table \ref{table:interpreting_single_BSM_results}. 

\subsubsection{Block level: \bsm{1}}

A block-level BSM (referred to \bsm{1}) is done by performing \bsm{0} on each PLS in the block. The sign of the block-level Bell state is determined by the majority vote of the signs of the \bsm{0} results. Its letter is determined by the parity of the number of \bsm{0} results with $\psi$ letter: $\phi$ ($\psi$) if the number is even (odd).

Since $m$ is odd, the sign of the block-level Bell state is always determinable. The letter is not determinable if at least one \bsm{0} fails, which we regard that the \bsm{1} fails.

\subsubsection{Logical level: \bsm{2}}

A logical-level BSM (referred as \bsm{2}) is done by performing \bsm{1} on each block. The sign of the logical-level Bell state is determined by the parity of the number of \bsm{1} results with minus sign: plus (minus) if the number is even (odd). Its letter is determined by the majority vote of the letters of the \bsm{1} results excluding the failed ones.

Again, the sign of the logical-level Bell state is always determinable. Its letter is not determinable if all the \bsm{1}s fail or the resulting block-level Bell states have the same number of both letters. We regard these cases as failure of \bsm{2}.

\subsection{Fault-tolerance of concatenated Bell-state measurement}
\label{subsec:fault_tolerance_of_cbsm}

Now, we investigate fault-tolerance of the unoptimized CBSM scheme suggested in the previous subsection. We argue that the physical-level and block-level repetitions contribute to suppressing logical errors and failures, respectively.

First, $Z$($X$)-errors in the logical level are suppressed by the majority vote at the block (logical) level. Remark that the sign (letter) of a logical-level Bell state is determined only by the signs (letters) of the Bell states of the lower levels, as described in Table \ref{table:unoptimized_cbsm_scheme}. $Z$-errors (sign flips) in the physical level can be corrected by the majority vote in the block level, so do not cause a logical-level $Z$-error with a high probability. Similarly, $X$-errors (letter flips) in the physical level can be corrected by the majority vote in the logical level, so also do not cause a logical-level $X$-error with a high probability. Since $Z$-errors are much more common than $X$-errors in the physical level ($p_x/p_z \lessapprox 10^{-3}$), we can infer that the physical-level repetition is crucial for fault-tolerance.

However, we cannot assure that the repetitions always suppress logical errors. Although $Z$-errors can be corrected by the physical-level repetition, the block-level repetition has a rather negative effect on it. Due to the error correction by the physical-level repetition, a block-level BSM result does not have a $Z$-error with a high probability. However, any single remained $Z$-error among the block-level BSM results can cause a $Z$-error in the logical level. Therefore, a large value of the size of the block-level repetition ($n$) leads to vulnerability of the CBSM to $Z$-errors. A similar logic applies to $X$-errors; the physical-level repetition has a negative effect on it.

Next, we consider failures in the logical level. As explained in the previous subsection, a \bsm{2} fails if all the \bsm{1}s fail or the results of the \bsm{1}s have the same number of both letters, and a \bsm{1} fails if any single \bsm{0} fails. The block-level repetition thus suppresses failures of the CBSM, whereas the physical-level repetition makes it vulnerable to failures.

In summary, ignoring $X$-errors which are much more uncommon than $Z$-errors and failures, the physical(block)-level repetition contributes to making the CBSM tolerant to $Z$-errors (failures) but vulnerable to failures ($Z$-errors). Despite these negative effects, we numerically show in Sec.~\ref{sec:numerical_calculations} that a high success probability are still achievable if the survival rate of photons is high enough and the amplitude of the coherent state is large enough.

\subsection{Improved hardware-efficient CBSM scheme}
\label{subsec:cbsm_scheme_with_optimized_cost}

\begin{figure*}[tb]
    \centering
    \includegraphics[width=0.8\textwidth]{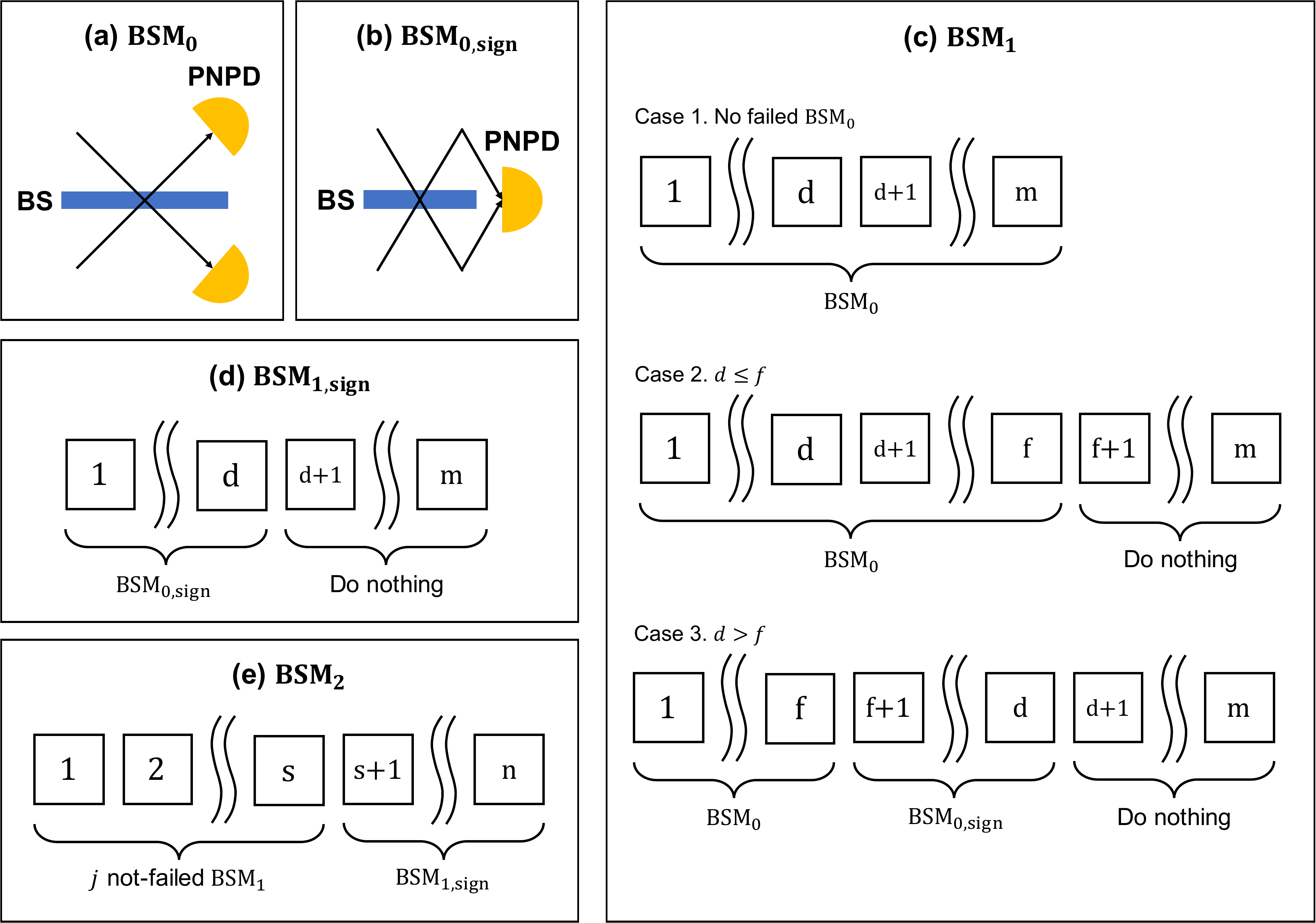}
    \caption{Overview of the hardware-efficient CBSM scheme. (a) For a physical-level BSM (\bsm{0}), a 50:50 beam splitter and two PNPDs are used. (b) For a physical-level partial BSM detecting only the sign (\bsms{0}), a single PNPD is necessary, instead of two. (c) For a block-level BSM (\bsm{1}), one of \bsm{0} and \bsms{0} is performed on each PLS one by one. We first define positive integers $d$ and $f$. $d$ is the index of the first PLS such that $\lceil m/2 \rceil$ of the physical-level BSM results until that PLS have the same sign. $f$ is the index of the first PLS such that the corresponding physical-level BSM fails, which is defined only if such a PLS exists. \textsc{(Case 1)} If there are no failed \bsm{0}s ($f$ is not defined), \bsm{0}s are performed on the entire PLSs. \textsc{(Case 2)} If $d \leq f$, \bsm{0}s are performed on the first $f$ PLSs and the remained PLSs are left untouched. \textsc{(Case 3)} If $d > f$, one performs \bsm{0}s for the first $f$ PLSs and \bsms{0}s for the next $d - f$ PLSs. The remained PLSs are left untouched. The reason to be able to do nothing for the last several PLSs in \textsc{Case 2} and \textsc{Case 3} is that these two cases correspond to the failure of the \bsm{1}, so more physical-level BSMs are meaningless if the sign of the block-level Bell state is determined. (d) For a block-level partial BSM detecting only the sign (\bsms{1}), \bsms{0}s are performed for the first $d$ PLSs and the remained PLSs are left untouched. (e) For a logical-level BSM (\bsm{2}), \bsm{1}s are performed one by one until $j$ not-failed \bsm{1} results are obtained, where $j$ is a controllable positive integer referred as the \textit{letter solidity parameter}. \bsms{1}s are then performed for the left blocks.}
    \label{fig:optimized_cbsm_scheme}
\end{figure*}
 
In this subsection, we suggest an improved CBSM scheme which is elaborately designed considering hardware efficiency. We explicitly define the \textit{cost} of a single trial of CBSM in the last part of this section, but we first regard it as the number of physical-level BSMs used for it. Note that the cost is generally not determined by the CBSM scheme alone; it can be different for each trial of CBSM.

The unoptimized scheme in Sec.~\ref{subsec:unoptimized_cbsm_scheme} always requires $nm$ physical-level BSMs, and here we suggest a way to decrease the number. The core idea is that it is redundant to perform `full' BSMs for all the PLSs or blocks, where the term `full' is used to emphasize that the BSM captures both sign and letter information of the Bell state. For some PLSs or blocks, it is enough to get only the sign ($\pm$) information of the Bell state or even do not measure it at all. Especially for the logical level, it is enough to perform full BSMs only for the first few blocks due to the biased noise. The hardware-efficient CBSM scheme which is presented from now on is summarized in Fig.~\ref{fig:optimized_cbsm_scheme}. 

\subsubsection{Physical level: \bsm{0} and \bsms{0}}

\bsm{0} is completely same with the scheme given in Sec.~\ref{sec:bsm_coherent_state_qubits}. Using a 50:50 beam splitter and two PNPDs (see Fig.~\ref{fig:optimized_cbsm_scheme}(a)), one of the four Bell states can be identified according to the results of the PNPDs, unless the two results are the same (failure). In the case of failure, only sign information of the Bell state can be captured.

However, we need another ingredient in the physical level for the hardware-efficient CBSM scheme: partial physical-level BSM identifying only the sign ($\pm$) of the physical-level Bell state, which we denote \bsms{0} (see Fig.~\ref{fig:optimized_cbsm_scheme}(b)). For \bsms{0}, one needs to measure the parity of $x + y$ in the Table \ref{table:interpreting_single_BSM_results}. Therefore, only one PNPD is needed for a \bsms{0} instead of two.

\subsubsection{Block level: \bsm{1} and \bsms{1}}

For a block-level BSM, we perform one of \bsm{0} or \bsms{0} on each PLS, one by one in order. The process is not parallel, since the determination between \bsm{0} and \bsms{0} is affected by the previous measurement results. We first define a positive integer $d \leq m$ by the index of the first PLS such that $\lceil m/2 \rceil$ of the physical-level BSM results until that PLS have the same sign. In other words, the result of the majority vote of the signs is already determined until $d$th physical-level BSM, and thus the sign information is no longer necessary. Also, we define a positive integer $f \leq m$ by the index of the first PLS such that the corresponding BSM fails, which is defined only if such a PLS exists.

Three cases are possible on \bsm{0}: no failed physical-level BSMs ($f$ is not defined), $d \leq f$, and $d > f$ (see Fig.~\ref{fig:optimized_cbsm_scheme}(c)). \textsc{(Case 1)} If there are no failed physical-level BSMs, it is same with the unoptimized scheme; \bsm{0}s are performed for all the PLSs. \textsc{(Case 2)} If $d \leq f$, \bsm{0}s are performed for the first $f$ PLSs. The remained $m - f$ PLSs are left untouched. \textsc{(Case 3)} If $d > f$, \bsm{0}s are performed for the first $f$ PLSs, and then \bsms{0}s are performed for the next $d - f$ PLSs. The remained $m - d$ PLSs are left untouched.

For all the three cases, the sign of the block-level Bell state is determined by the signs of the first $d$ \bsm{0} (or \bsms{0}) results. However, the letter is determined only for the first case by the parity of the number of \bsm{0} results with letter $\psi$. For the second and third case, there exists a failed \bsm{0}, so the number of results with letter $\psi$ is ambiguous. Hence, the \bsm{1} fails in these two cases. This is the reason to be able to do nothing on the last several PLSs after the sign of the block-level BSM is determined.

Like the physical level, we also consider partial block-level BSM which determines only the sign of the block-level Bell state (\bsms{1}) (see Fig.~\ref{fig:optimized_cbsm_scheme}(d)). For \bsms{1}, \bsms{0}s are performed for the first $d$ PLSs, and the remained PLSs are left untouched. The sign of the block-level Bell state is determined by the majority vote of the results of the first $d$ \bsms{0} results.

\subsubsection{Logical level: \bsm{2}}

For a logical-level BSM (\bsm{2}) (see Fig.~\ref{fig:optimized_cbsm_scheme}(e)), \bsm{1}s are performed one by one until we get $j$ not-failed results. $j$ is a controllable positive integer referred as the \textit{letter solidity parameter} which means that high values of $j$ lead to high probabilities to get correct letter information. After that, \bsms{1}s are performed for the remained blocks.

The sign of the resulting logical-level Bell state is determined by the parity of the number of \bsm{1} or \bsms{1} results with minus sign. The letter is determined by the majority vote of the letters among the first $j$ not-failed \bsm{1} results.

Note the difference between \bsm{2} and \bsm{1}: For \bsm{2}, the majority vote is taken for the first $j$ not-failed \bsm{1}s with a fixed $j$, while for \bsm{1}, the majority vote is taken when the result of the majority vote on the total PLSs is definitely determined. This asymmetry comes from the fact that the noise is strongly biased; $X$-errors are much less likely to occur compared to $Z$-errors in \bsm{0} as shown in Fig.~\ref{fig:phys_bsm_probs}. Therefore, when taking the majority vote of the letters of the \bsm{2} results, it is enough to use only a few \bsm{1} results to correct $X$-errors. On the other hands, the majority vote of the signs of the physical-level BSM results should be taken for a large number of PLSs.

\subsubsection{Calculation of the cost}
\label{subsubsec:calculation_of_cost}

At the beginning of this subsection, we regard the cost of a single CBSM by the number of physical-level BSMs used for the measurement. However, considering that PNPDs are the most difficult elements when implementing the \bsm{0} scheme and a \bsms{0} uses one of them while a \bsm{0} uses two, it is reasonable to assign each \bsms{0} half the cost of one \bsm{0}.
\begin{definition}
\label{def:cost_function}
The \textit{cost function} $C$ of a single trial of CBSM is defined by
\begin{align}
    C := N_{\mathrm{BSM}_0} + \frac{1}{2} N_{\mathrm{BSM}_{0}^{\mathrm{sign}}}, \label{eq:cost_function_def}
\end{align}
where $N_{\text{BSM}_{0}}$ and $N_{\mathrm{BSM}_{0}^{\mathrm{sign}}}$ are the number of \bsm{0}s and \bsms{0}s used for the CBSM, respectively. Also, we define the \textit{expected cost} $C_{exp}(n, m, \alpha, j; \eta)$ by the expectation value of the cost $C$ for the CBSM scheme specified by the parameters $(n, m, \alpha, j)$ and the photon survival rate $\eta$, with the assumption that the initial state before suffering photon loss is one of the four logical Bell states with equal probabilities.
\end{definition}

We use the expected cost $C_{exp}$ as a measure of hardware-efficiency of a CBSM scheme. It is straightforward to see that the CBSM scheme in the previous subsection has a less expected cost than the unoptimized one in Sec.~\ref{subsec:unoptimized_cbsm_scheme}. Not only that, it is designed to minimize the expected cost. For \bsm{1}, the numbers of \bsm{0} and \bsms{0} are minimized while keeping the result to be the same with that of the corresponding \bsm{1} in the unoptimized scheme. For \bsm{2}, the expected cost is determined by the controllable letter solidity parameter $j$.

\subsection{Parallelization of concatenated Bell-state measurement}
\label{subsec:distributed_nature_of_cbsm}

The two CBSM schemes in Sec. \ref{subsec:unoptimized_cbsm_scheme} and \ref{subsec:cbsm_scheme_with_optimized_cost} are processed in a completely or partially distributed manner, which makes efficient information processing possible by parallelization. The unoptimized scheme is done in a completely distributed manner, i.e., a \bsm{2} is split by $2nm$ \bsm{0}s, each of which is performed independently. The \bsm{0} results are collected classically to deduct the logical-level BSM result.

The hardware-efficient scheme also can be done in a partially distributed manner allowing partial parallelization, with requirements of classical communication channels between different PLSs and blocks. In a \bsm{2}, \bsm{1}s can be done parallelly for the first $j$ blocks, then one by one until obtaining $j$ not-failed \bsm{1} results, where $j$ is the letter solidity parameter. \bsms{1}s for the remained blocks also can be done parallelly. In a \bsm{1}, \bsm{0}s should be done one by one until a \bsm{0} fails, so \bsm{0}s in all the three cases cannot be done parallelly. \textsc{Case 3} can be partially parallelized only if $f < m/2$: \bsms{0}s can be done parallelly for $(f+1)$th to $\lceil m/2 \rceil$th PLS since $d$ is always larger than $m/2$. In \bsms{1}, \bsms{0} can be done parallelly for the first $\lceil m/2 \rceil$ PLSs, then one by one for the remained PLSs.

Therefore, the hardware efficiency is the result of the sacrifice of parallelization. We can still widen the range of parallelization by adjusting the scheme appropriately at the expense of reducing hardware efficiency. For example, in a \bsm{2}, \bsm{1}s can be done for the first $j$ blocks, not for the first not-failed $j$ blocks. Moreover, in a \bsm{1} and \bsms{1}, instead of determining the type of BSM (\bsm{0} or \bsms{0}) separately for each PLS, we can divide the PLSs into several groups and perform BSMs with the same type parallelly on PLSs in each group. However, we use the original hardware-efficient CBSM scheme for the numerical simulation in Sec.~\ref{sec:numerical_calculations} to figure out the best possible performance.

\section{Probability distributions of concatenated Bell-state measurement results}
\label{sec:prob_dist_of_measurement_results}

In this section, we present the analytic expressions of the probability distributions of CBSM results conditioning to the initial Bell state before suffering photon loss. We only consider the unoptimized CBSM scheme, since the measurement results of the hardware-efficient CBSM scheme is the direct consequence of those of the unoptimized scheme. Here, we show only the final results. A brief outline for inducing the results is presented in Appendix \ref{app:method_sampling_measurement_results}.

The results of this section have two important meanings. First, the probability distributions are written in simple matrix-form expressions, which makes it possible to sample arbitrary CBSM results at a high rate, since a matrix calculation can be done much faster on a computer compared to calculating the same thing by simple loops. Second, the results can be easily generalized to any CBSM schemes with other encoding methods such as multi-photon polarization encoding \cite{lee2019fundamental}.

\subsection{Probability distributions of block-level results}
\label{subsec:prob_dist_of_bsm_1_results}

We first find the probability distributions of block-level BSM results, conditioning to the initial block-level Bell state. A single \bsm{1} result can be expressed by two vectors $\vb{x}, \vb{y} \in \{0, 1, 2, 3 \}^m$, where the $i$th elements of them are the two PNPD results of the $i$th PLS. What we want is the conditional probability $\P(\vb{x}, \vb{y} | B_1)$ for $\ket{B_1} \in \mathcal{B}_1 := \qty{ \ket{\phi^{(m)}_\pm}, \ket{\psi^{(m)}_\pm} }$.

First, we define $4\times4$ matrices $\vb{\tilde{M}}^\pm_{x,y}$ for $x, y \in \qty{0, 1, 2, 3 }$ as:
\begin{align*}
    \vb{\tilde{M}}^\pm_{x,y} :=
    \begin{pmatrix}
        M^{\pm}_{11} & M^{\pm}_{12} & M^{\pm}_{12} & M^{\pm}_{22} \\
        M^{\pm}_{12} & M^{\pm}_{11} & M^{\pm}_{22} & M^{\pm}_{12} \\
        M^{\pm}_{12} & M^{\pm}_{22} & M^{\pm}_{11} & M^{\pm}_{12} \\
        M^{\pm}_{22} & M^{\pm}_{12} & M^{\pm}_{12} & M^{\pm}_{11}
    \end{pmatrix},
\end{align*}
where
\begin{align*}
    M^{\pm}_{11} &:= \bra{\phi_\pm} \hat{M}_{x, y} \ket{\phi_\pm}, \\
    M^{\pm}_{12} &:= \bra{\phi_\pm} \hat{M}_{x, y} \ket{\psi_\pm}, \\
    M^{\pm}_{22} &:= \bra{\psi_\pm} \hat{M}_{x, y} \ket{\psi_\pm}
\end{align*}
are the matrix elements of POVM elements of \bsm{0} and can be calculated from Eqs.~\eqref{eqs:M_xy} in Appendix \ref{app:povm_elements}. 
The conditional probability $\P(\vb{x}, \vb{y} | B_1)$, where the $k$th element of $\vb{x}$($\vb{y}$) is $x_k$($y_k$), is then:
\begin{subequations}
\label{eqs:block_cond_probs_by_v_result}
\begin{align}
    \P(\vb{x}, \vb{y}|\phi^{(m)}_\pm) &= \frac{1}{2} \tilde{N}_{\pm}(1, m)^2  v^\pm_{m1} (\vb{x}, \vb{y}), \\
    \P(\vb{x}, \vb{y}|\psi^{(m)}_\pm) &= \frac{1}{2} \tilde{N}_{\pm}(1, m)^2 v^\pm_{m4} (\vb{x}, \vb{y}),
\end{align}
\end{subequations}
where $\tilde{N}_\pm (1, m)$ is defined in Eq.~\eqref{eq:N_tilde_def} and $v^\pm_{m\mu} (\vb{x}, \vb{y})$ is the $\mu$th element of a four-dimensional vector $\vb{v}^\pm_m (\vb{x}, \vb{y}) = \vb{\tilde{M}}^\pm_{x_m, y_m} \cdots \vb{\tilde{M}}^\pm_{x_1, y_1} (1,~0,~0,~0)^T$.

A brief outline for inducing these results is presented in Appendix \ref{subapp:prob_dist_of_bsm_1_results}.

\subsection{Probability distributions of logical-level results}
\label{subsec:prob_dist_of_bsm_2_results}

Now, we consider the probability distributions of logical-level results conditioning to the initial logical-level Bell state, which is the goal of this section. A single CBSM result can be expressed by two matrices $\vb{X}, \vb{Y} \in \{ 0, 1, 2, 3 \}^{n \times m}$, where the $(i,k)$ elements of them are the two PNPD results of the $k$th PLS of the $i$th block. What we want is the conditional probability $\P(\vb{X}, \vb{Y} | B_2)$ for $\ket{B_2} \in \mathcal{B}_2 := \qty{ \ket{\Phi_\pm}, \ket{\Psi_\pm} }$.

We first define $2 \times 2$ matrices $\vb{\tilde{L}}^{\phi}_{\vb{x},\vb{y}}$ and $\vb{\tilde{L}}^{\psi}_{\vb{x},\vb{y}}$ where $\vb{x}, \vb{y} \in \qty{ 0, 1, 2, 3 }^m$ in the similar way with the block-level case:
\begin{align*}
    \vb{\tilde{L}}^{\phi(\psi)}_{\vb{x}, \vb{y}} :=
        \begin{pmatrix}
            L^{\phi(\psi)}_{+} & L^{\phi(\psi)}_{-} \\
            L^{\phi(\psi)}_{-} & L^{\phi(\psi)}_{+}
        \end{pmatrix},
\end{align*}
where
\begin{align}
\label{eq:L_def_result}
    &L^{\phi(\psi)}_\pm := \qty[ 1 \pm u(\alpha, m)^2 ] \nonumber \\
    &\quad \times \expval{ \bigotimes_{k=1}^m \hat{M}_{x_k, y_k} }{ \phi^{(m)}_\pm \qty( \psi^{(m)}_\pm ) },
\end{align}
$u(\alpha, m)$ is defined in Eq.~\eqref{eq:u_def}, and $x_k$($y_k$) is the $k$th element of $\vb{x}$($\vb{y}$). We note that the RHS of Eq.~\eqref{eq:L_def_result} can be calculated from Eqs.~\eqref{eqs:block_cond_probs_by_v_result}. The conditional probability $\P(\vb{X}, \vb{Y} | B_2)$, where the $i$th row vector of $\vb{X}$($\vb{Y}$) is $\vb{x}_i$($\vb{y}_i$), is then
\begin{align*}
    \P(\vb{X}, \vb{Y}|\Phi_+ (\Psi_+)) &= \tilde{N}_{+}(n, m)^2 w^{\phi (\psi)}_{n1} (\vb{X}, \vb{Y}), \\
    \P(\vb{X}, \vb{Y}|\Phi_- (\Psi_-)) &= \tilde{N}_{-}(n, m)^2 w^{\phi (\psi)}_{n2} (\vb{X}, \vb{Y}), 
\end{align*}
where $\tilde{N}_\pm (n, m)$ is defined in Eq.~\eqref{eq:N_tilde_def} and $w_{n\mu}^{\phi(\psi)} (\vb{X}, \vb{Y})$ is the $\mu$th element of the two-dimensional vector $\vb{w}^{\phi(\psi)}_n (\vb{X}, \vb{Y}) := \vb{\tilde{L}}^{\phi(\psi)}_{\vb{x}_n, \vb{y}_n} \cdots \vb{\tilde{L}}^{\phi(\psi)}_{\vb{x}_1, \vb{y}_1} (1, 0)^T$. A brief outline for inducing these results is presented in Appendix \ref{subapp:prob_dist_of_bsm_2_results}.

In conclusion, one can calculate the probability distributions of CBSM results by systematical matrix operations as described in this and the previous subsection. The probability distributions then can be used to sample the CBSM results for numerical calculations.

\section{Numerical calculations}
\label{sec:numerical_calculations}

In this section, we show the results of the numerical calculations. We use the Monte-Carlo method for the simulation: sampling the measurement results randomly and counting the number of successes, errors, and failures. We sample the result of each physical-level BSM one by one in order, which is exponentially faster than sampling the entire measurement results at once. The detailed method for sampling the CBSM results using the results of Sec.~\ref{sec:prob_dist_of_measurement_results} is presented in Appendix \ref{app:method_sampling_measurement_results}.

Remark that there are four free parameters related to the hardware-efficient CBSM scheme: $n$, $m$, $\alpha$, and $j$. $n$ and $m$ determine the block-level and physical-level repetition size of the scheme, respectively. $\alpha$ is the amplitude of the coherent state constituting the logical basis. $j$ is the letter solidity parameter which is the number of not-failed blocks used for the majority vote of letters in \bsm{2}.

\subsection{Performance analysis}
\label{subsec:performance_analysis}

Now, we analyze the performance of the hardware-efficient CBSM scheme suggested in Sec.~\ref{subsec:cbsm_scheme_with_optimized_cost} by calculating numerically the success, error, and failure probabilities of the scheme with various settings of the parameters $(n, m, \alpha, j)$. 
For the simulation, we assume that both systems have the same photon survival rates $\eta$. 
We use the Monte-Carlo method as mentioned before. 
For each trial, we first choose one of the four Bell states as the initial state with equal probabilities, sample the physical-level BSM results with respect to the selected initial state, and determine the logical Bell state by the hardware-efficient CBSM scheme. 
Repeating this trials many times, we determine the success ($p_i$), $Z$-error ($p_z$), and failure probabilities ($p_{fail}$) of the CBSM scheme. 
We also calculate the expected cost $C_{exp}$ defined in \textit{Definition~\ref{def:cost_function}}.

\begin{figure}[tb]
    \centering
    \includegraphics[width=\linewidth]{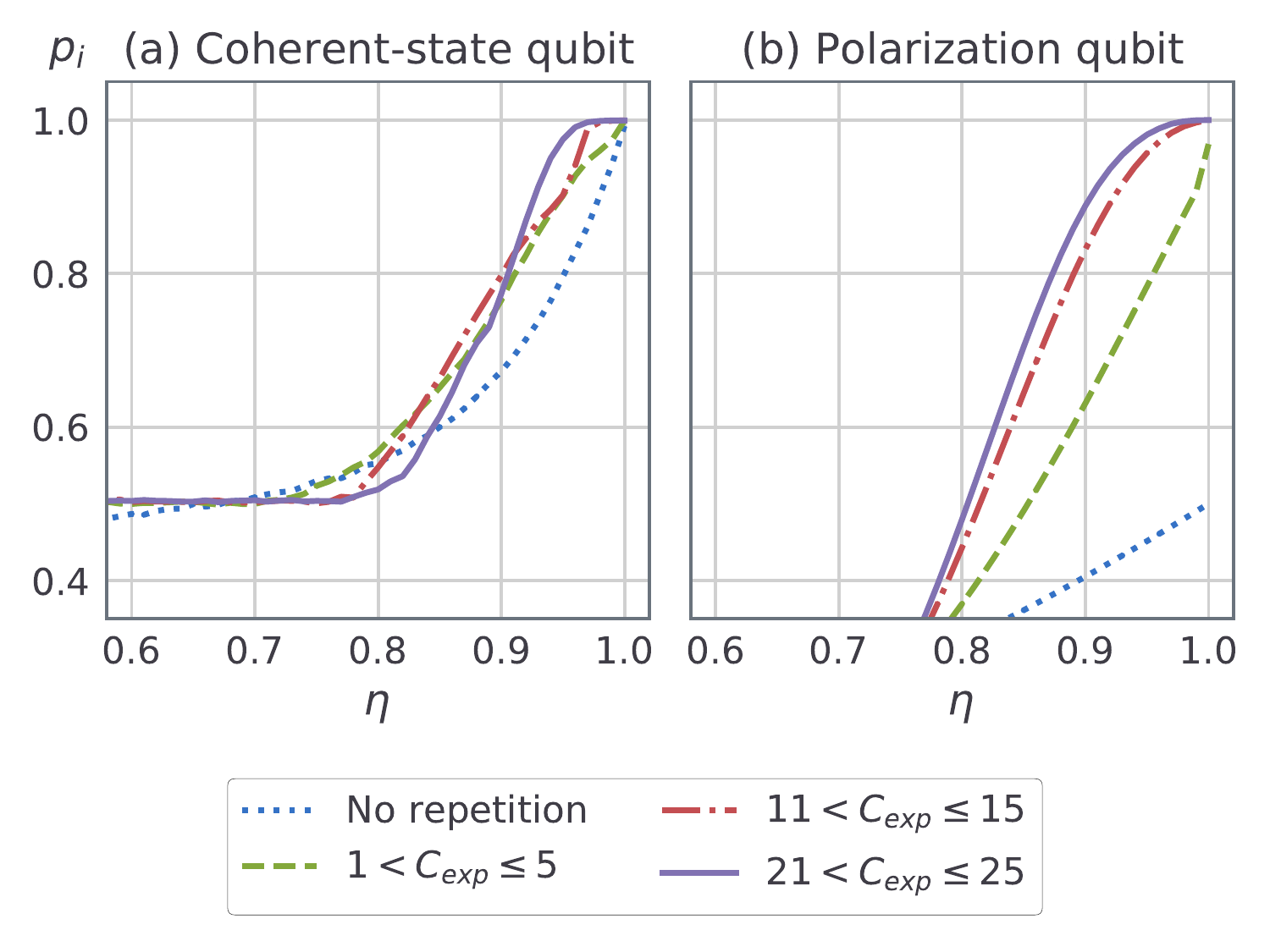}
    \caption{Success probabilities $p_i$ of CBSM with (a) coherent-state qubits and (b) multi-photon polarization qubits against the photon survival rate $\eta$ for different ranges of the expected cost $C_{exp}$. For coherent-state qubits, the amplitude $\alpha$ is fixed to $\alpha=1.6$ and the letter solidity parameter $j$ is chosen to maximize $p_i$ for each $\eta$ and range of $C_{exp}$. For polarization qubits, we follow the CBSM scheme proposed in \cite{lee2019fundamental}. In this case, we define $C_{exp} := nm$, which is the number of physical-level BSMs used for one CBSM. (a) shows that the repetition indeed contributes to enhance the success probability. Comparing (a) and (b), we can see that the CBSM with coherent-state qubits outperforms that with polarization qubits when the repetition size is relatively small.}
    \label{fig:succ_rate_against_eta}
\end{figure}

Figure~\ref{fig:succ_rate_against_eta} illustrates the success probability $p_i$ of CBSM with coherent-state qubits and polarization qubits \cite{lee2019fundamental} against the photon survival rate $\eta$ for different ranges of the expected cost $C_{exp}$, where $p_i$ is maximized for each $\eta$ and $C_{exp}$. 
Figure~\ref{fig:succ_rate_against_eta}(a) shows that the repetition indeed enhances the performance  if $\eta \gtrapprox 0.8$ compared to the case without repetition. 
The effect of the repetition is especially crucial if $\eta$ is close to unity. 
For example, if $\eta=0.95$, $p_i = 0.80$ without repetition, but it reaches 0.90 with just a little repetition ($C_{exp} \leq 5$), and up to 0.99 for $31 < C_{exp} \leq 35$. 
In other words, it is the clear evidence that high success rates close to unity are achievable by CBSM if the photon survival rate is sufficiently high. 
Meanwhile, comparing Fig.~\ref{fig:succ_rate_against_eta}(a) and (b), we can see that the CBSM with coherent-state qubits outperforms that with multi-photon polarization qubits when the repetition size is relatively small ($C_{exp} \leq 5$). 
For instance, if $\eta=0.99$, the CBSM with coherent-state qubits achieves $p_i = 0.90$ for $C_{exp} \leq 5$, while that with multi-photon polarization qubits reaches only $p_i = 0.78$.

\begin{figure}[tb]
    \centering
    \includegraphics[width=\linewidth]{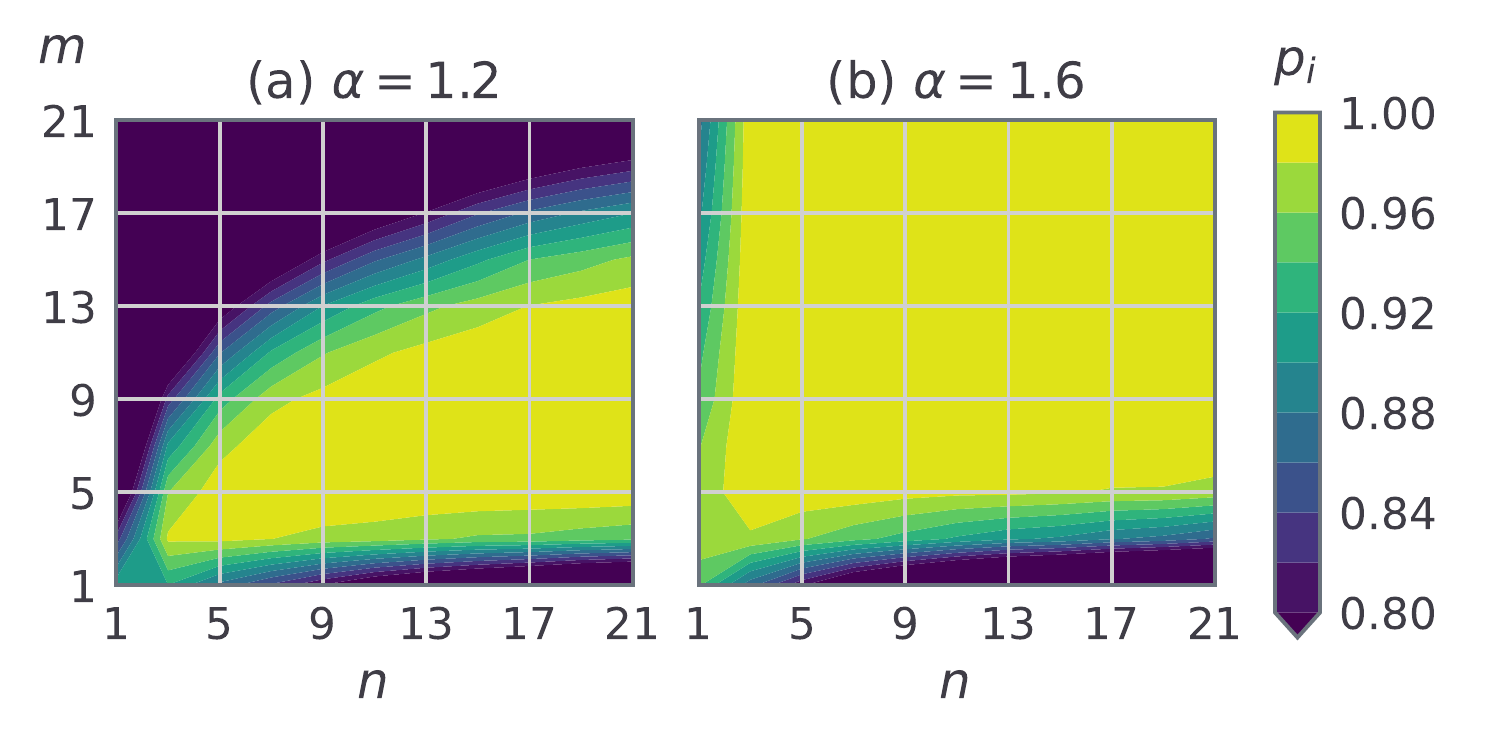}
    \includegraphics[width=\linewidth]{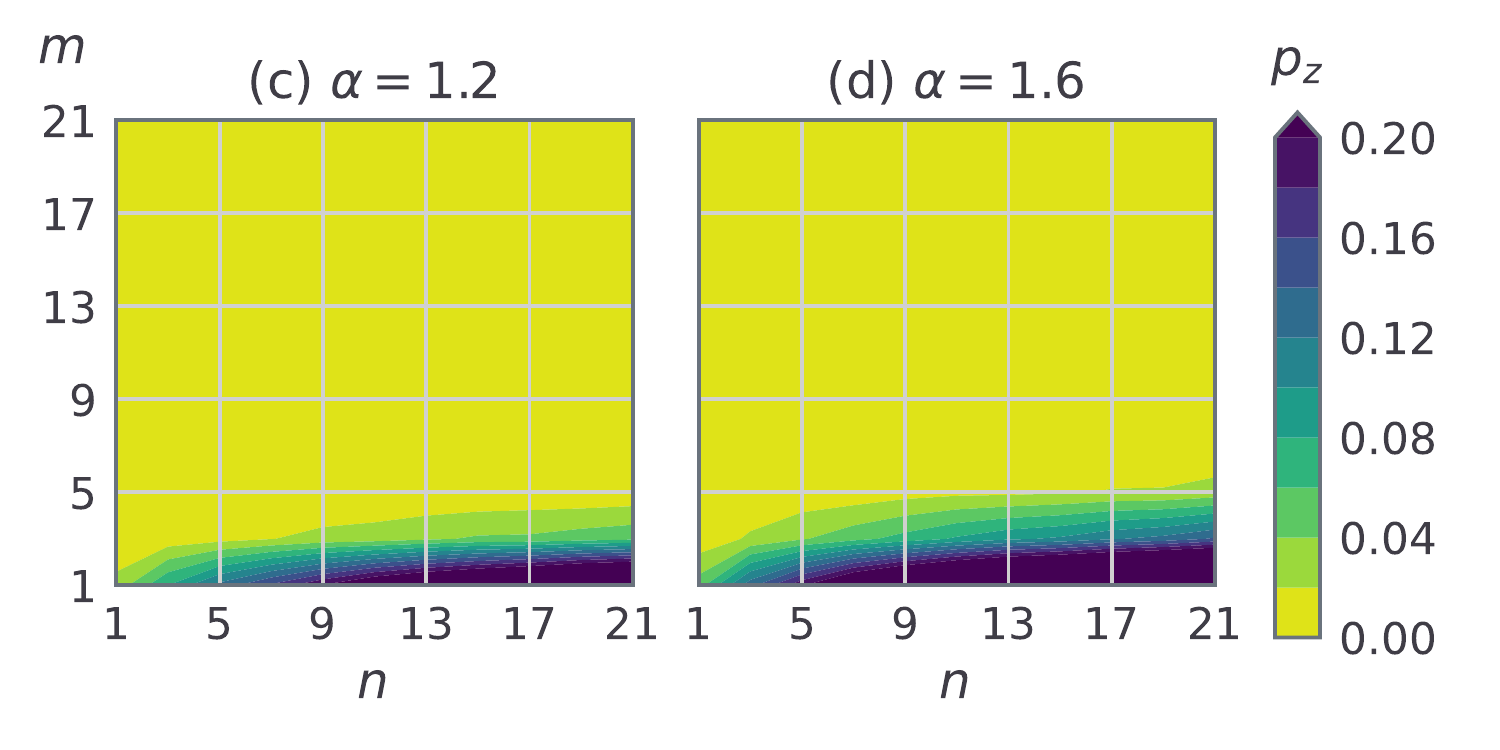}
    \includegraphics[width=\linewidth]{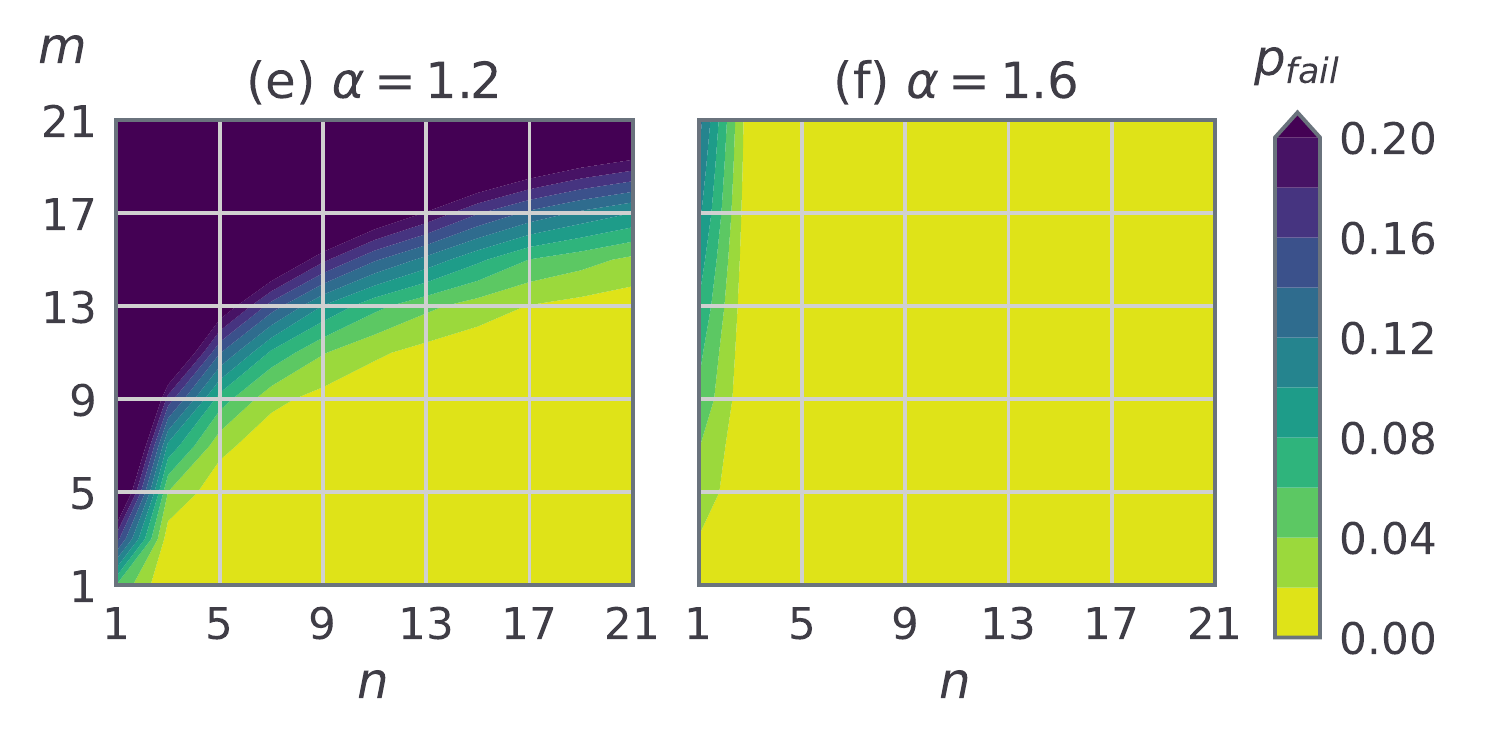}
    \caption{Success ($p_i$), $Z$-error ($p_z$), and failure probabilities $p_{fail}$ of CBSM against the repetition sizes $n$ and $m$ for coherent-state amplitudes $\alpha=1.2$ and $\alpha=1.6$. The photon survival rate $\eta$ is fixed to 0.99, and $j$ is selected to maximize $p_i$ for each $(n,m)$ point. It clearly shows that physical-level repetition suppresses $Z$-error and block-level repetition suppress failure.}
    \label{fig:rates_against_n_m}
\end{figure}

In Fig.~\ref{fig:rates_against_n_m}, we compares the success ($p_i$), $Z$-error ($p_z$), and failure probabilities ($p_{fail}$) against $n$ and $m$, for two different values of $\alpha$: $\alpha = 1.2$ and 1.6. 
From this figure, we can check the dependence of $p_z$ and $p_{fail}$ on the repetition; it clearly shows that the physical-level repetition ($m > 1$) suppresses $Z$-errors and the block-level repetition ($n > 1$) suppresses failures, as argued in Sec.~\ref{subsec:fault_tolerance_of_cbsm}. 
Moreover, the negative effects discussed in Sec.~\ref{subsec:fault_tolerance_of_cbsm} that the physical(block)-level repetition makes the CBSM vulnerable to failures ($Z$-errors) are also shown in the figure, and in spite of them, the success probability close to unity still can be achieved.

\begin{figure*}[tb]
    \centering
    \includegraphics[width=\textwidth]{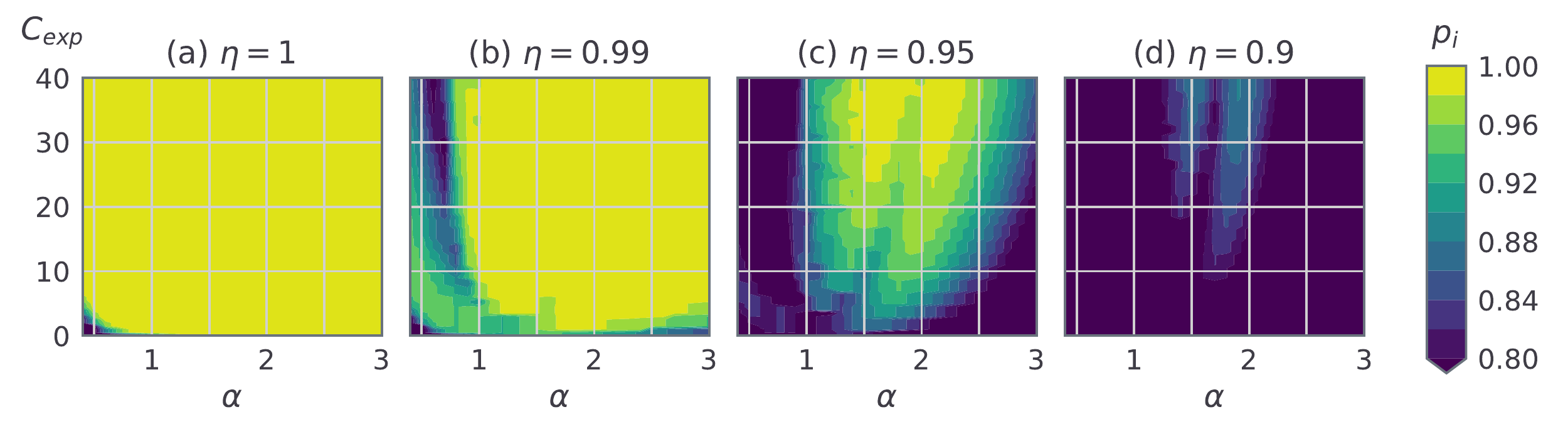}
    \caption{Success probabilities $p_i$ against the coherent-state amplitude $\alpha$ and the expected cost $C_{exp}$ for (a) $\eta=1$, (b) $\eta=0.99$, (c) $\eta=0.95$, and (d) $\eta=0.9$, where $\eta$ is the photon survival rate of both parties. $p_i$ is selected by $\max \qty{ p_i (n, m, \alpha_0, j) \middle| C_{exp}(n, m, \alpha_0, j; \eta) \in [C_0 - 2, C_0 + 2) }$ for each point $(\alpha_0, C_0)$. The figure indicates that a large value of $\alpha$ does not always guarantee a high success probability, which is especially evident in (c).}
    \label{fig:succ_rate_against_alpha_cost}
\end{figure*}

Lastly, the success probability $p_i$ against $\alpha$ and the expected cost $C_{exp}$ for four different survival rates ($\eta=1$, 0.99, 0.95, and 0.9) is plotted in Fig.~\ref{fig:succ_rate_against_alpha_cost}. 
The figure shows that the success probability over $0.98$ can be reached for $\eta \geq 0.95$ and appropriate values of $\alpha$, if sufficiently large costs of the CBSM is available. 
In lossless case ($\eta=1$), the success probability reaches very close to unity for any $\alpha \gtrapprox 0.4$ with just a little repetition. 
As the photon survival rate gets smaller, appropriately large values of $\alpha$ and cost are required for reaching high success probabilities. 
In detail, to reach $p_i > 0.98$, we need $\alpha \gtrapprox 0.8$ for $\eta=0.99$ and $\alpha \gtrapprox 1.4$ for $\eta=0.95$. 
Nonetheless, the figure also indicates that a higher value of $\alpha$ does not always guarantee a higher success rate due to dephasing by photon loss, which is especially evident in (c) $\eta=0.95$.

\subsection{Quantum repeater with concatenated Bell-state measurement}
\label{subsec:quantum_repeater}

\begin{figure*}[tb]
    \centering
    \includegraphics[width=\textwidth]{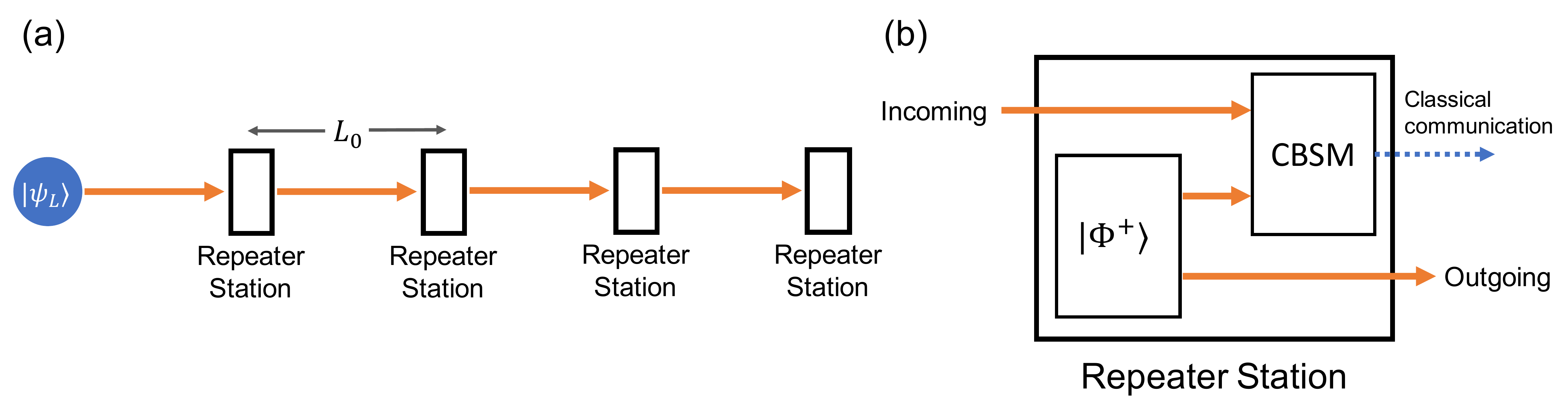}
    \caption{(a) Schematic of quantum information transmission through the quantum repeater scheme. Quantum information encoded in modified parity encoding is transmitted to the other end. It passes through multiple repeater stations where the interval is $L_0$. (b) Schematic of processes inside a repeater station. A Bell state $\ket{\Phi^+}$ is prepared inside the station and a CBSM is performed between the incoming qubit and one side of the Bell state. The quantum information inside the incoming qubit is then teleported to the other side of the Bell state, which is then transmitted to the next repeater station. The measurement result of the CBSM is sent classically to the final end for recovering the original quantum information. Due to fault-tolerance of the CBSM scheme, each repeater station can correct possible logical errors from photon loss which the incoming qubit suffers.}
    \label{fig:repeater_schematic}
\end{figure*}

\begin{figure*}[tb]
    \centering
    \includegraphics[width=\textwidth]{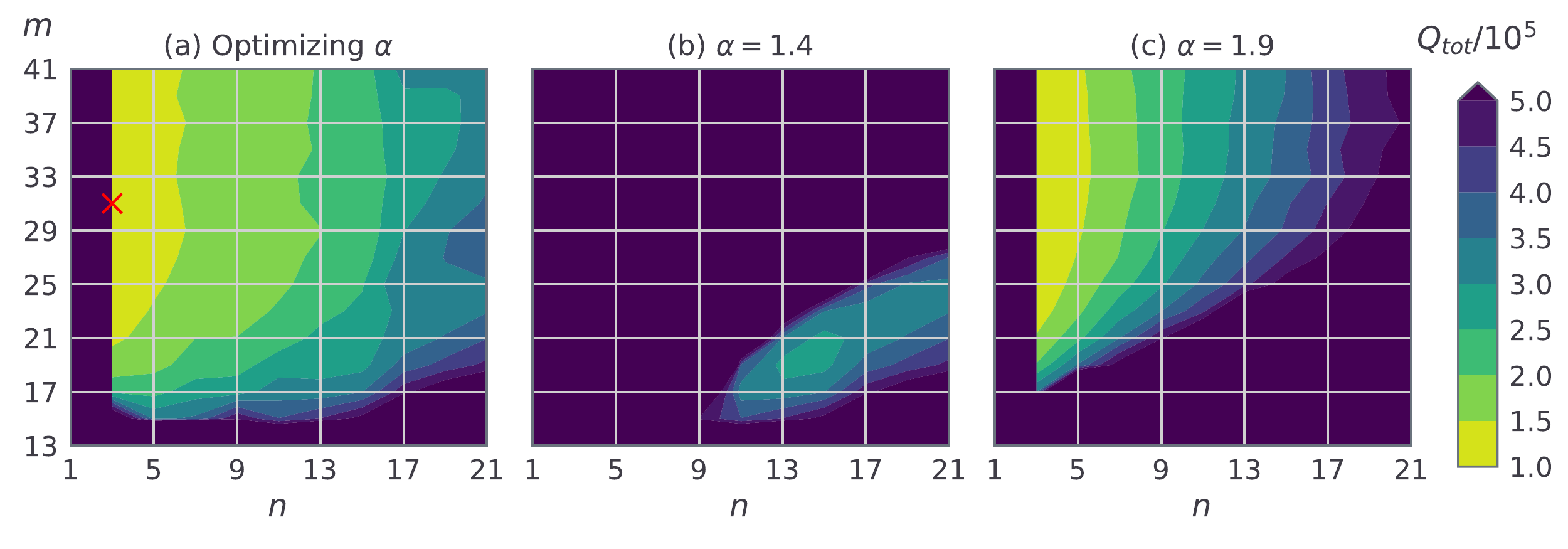}
    \includegraphics[width=\textwidth]{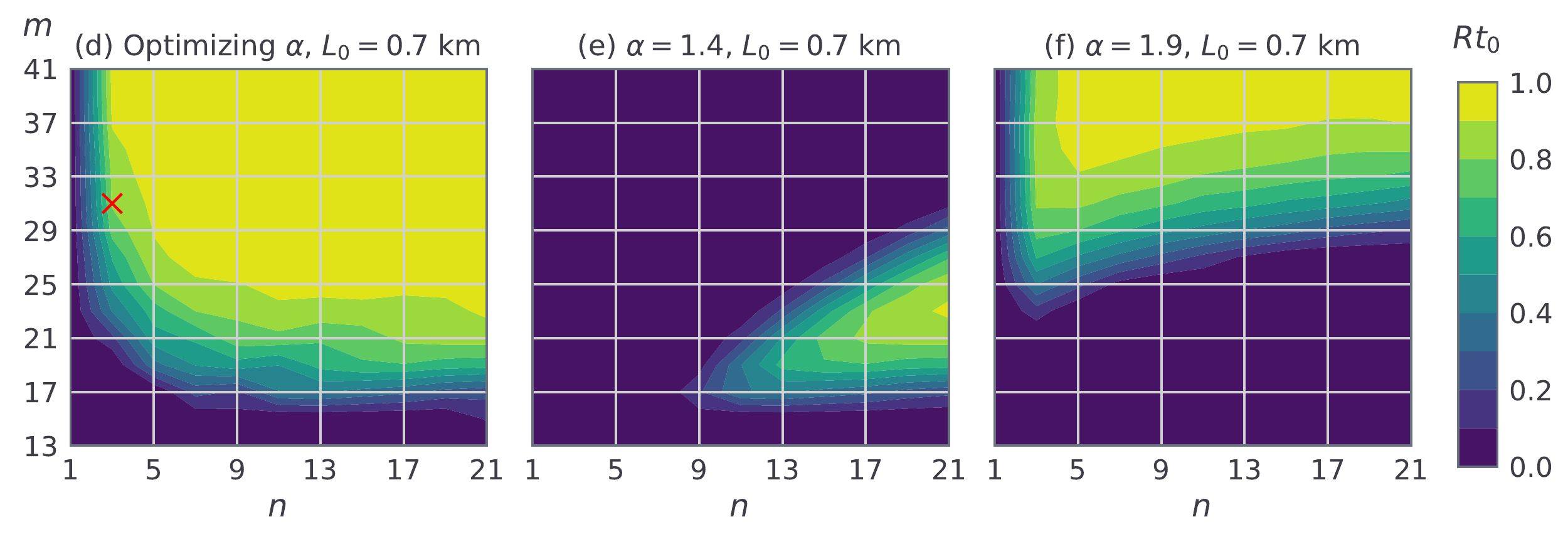}
    \caption{(top) Effective total cost $Q_{tot}$ and (bottom) expected key length $Rt_0$ of the quantum repeater against repetition sizes $n$ and $m$, for three different settings of the coherent-state amplitude $\alpha$: optimizing $\alpha$, fixing $\alpha=1.4$, and fixing $\alpha=1.9$. We fix the total distance $L=1000$ km and the internal photon survival rate in each station $\eta_0=0.99$. For calculating $Rt_0$, we also fix the station interval $L_0=0.7$ km. For each $(n,m)$ point, other parameters such as the letter solidity parameter $j$ and the station interval $L_0$ (only for $Q_{tot}$) are selected to minimize $Q_{tot}$ or maximize $Rt_0$. The 'X' marks in (a) and (d) indicate the optimal point where $Q_{tot}$ is minimized. The parameters at this point are $(n, m, \alpha, j) = (3, 31, 1.9, 1)$ and $L_0 = 0.7 \text{ km}$. $Rt_0 = 0.71 \pm 0.02$ and $Q_{tot} = (1.019 \pm 0.003) \times 10^5$ at this point, where the range is the 95\% confidence interval.}
    \label{fig:eff_total_cost_and_Rt0_against_n_m}
\end{figure*}

\begin{figure}[tb]
    \centering
    \includegraphics[width=\linewidth]{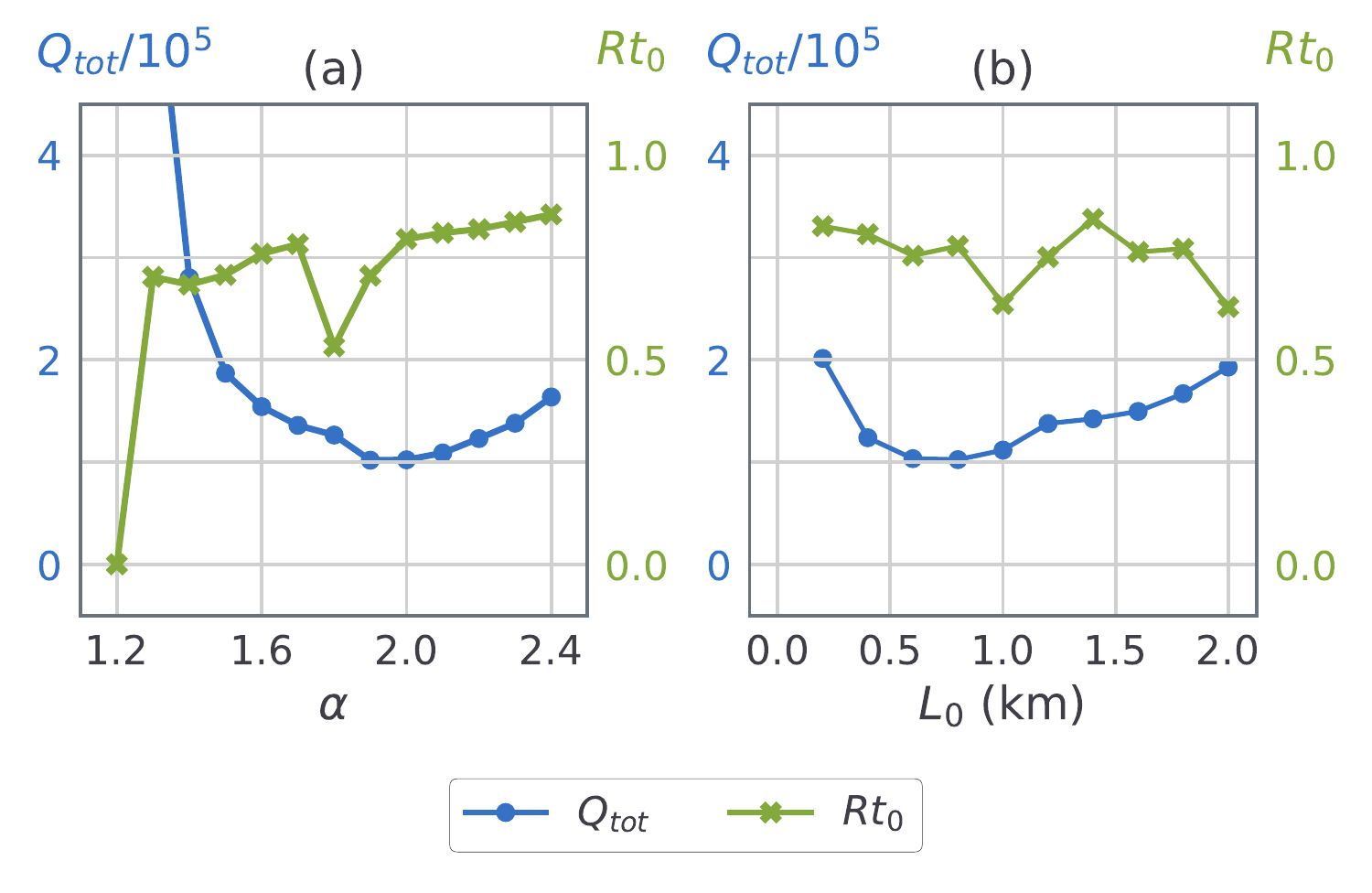}
    \caption{Optimal effective total cost $Q_{tot}$ and the corresponding expected key length $Rt_0$ against (a) the coherent-state amplitude $\alpha$ and (b) the repeater station interval $L_0$. For each point, other parameters ($n$, $m$, $j$, $L_0$ for (a) / $n$, $m$, $j$, $\alpha$ for (b)) are selected to minimize $Q_{tot}$, and the value of $Rt_0$ corresponds to that optimal set of parameters. Overall, $Q_{tot}$ is minimum at $\alpha=1.9$ and $L_0=0.7$ km.}
    \label{fig:eff_total_cost_Rt0_against_alpha_L0}
\end{figure}

In this subsection, we investigate the performance of the quantum repeater scheme which uses the suggested CBSM scheme for quantum error correction, as one of the key applications of BSM.

\subsubsection{Network design}

We follow the network design in Ref. \cite{lee2019fundamental}, which suggests an all-optical quantum network with quantum repeater exploiting the CBSM scheme with multi-photon polarization qubits. As shown in Fig.~\ref{fig:repeater_schematic}(a), we consider one-way quantum communication with which a qubit encoded by the modified parity encoding is transmitted to the other end. While traveling between two ends with the total distance of $L$, the qubit passes through multiple repeater stations separated by intervals of $L_0$. Figure~\ref{fig:repeater_schematic}(b) illustrates the processes inside each repeater station. In each of them, a Bell state $\ket{\Phi_+}$ is prepared and a CBSM is performed jointly on the incoming qubit and one side of the Bell state. The quantum information in the incoming qubit is then teleported to the other side of the Bell state, which is transmitted to the next station. The measurement result of the CBSM in each station is sent classically to the final end for recovering the original quantum information. Because of fault-tolerance of the CBSM scheme, each repeater station can correct possible logical errors originated from photon loss, which makes a long-range transmission of quantum information possible.

We assume two sources of photon loss: internal loss in each repeater station and loss during transmission between stations with survival rates of $\eta_0$ and $\eta_{L_0} := e^{-L_0/L_{\mathrm{att}}}$, respectively, where $L_{\mathrm{att}} = 22~\text{km}$ is the attenuation length. Therefore, the survival rates of two systems on which CBSM is jointly performed is $\eta_1 := \eta_0 e^{-L_0/L_{att}}$ and $\eta_2 := \eta_0$.

\subsubsection{Quantification of the performance}

One way to quantify the performance of a quantum repeater scheme is the \textit{asymptotic key generation rate} $R$ of quantum key distribution (QKD), which is the expected length of a fully secure key that can be produced per unit time \cite{scarani2009security, muralidharan2014ultrafast}. More precisely, it is the product of the raw-key rate, which is the length of a raw key that can be produced per unit time, and the secret fraction, which is the fraction of the length of a fully secure key to the length of a raw key in the asymptotic case of $N \rightarrow \infty$ where $N$ is the number of signals \cite{scarani2009security}. We use $Rt_0$ as the measure of performance where $t_0$ is the time taken in one repeater station, which we call the \textit{expected key length}. The expected key length is given by \cite{muralidharan2014ultrafast}:
\begin{equation}
    Rt_0 = \max \qty[ P_s \qty{ 1 - 2h(Q) }, 0 ], \label{eq:R_def}
\end{equation}
where $P_s$ is the probability not to fail during the entire transmission, $Q$ is the average quantum bit error rate (QBER), and $h(Q) := -Q \log_2(Q) - (1 - Q) \log_2(1 - Q)$ is the binary entropy function. The probability $P_s$ is given by:
\begin{equation*}
    P_s = (1 - p_{fail})^{L/L_0},
\end{equation*}
where $p_{fail}$ is the failure probability of a CBSM in a single repeater station. The average QBER $Q$ is defined by $Q = (Q_X + Q_X)/2$, where $Q_X$ and $Q_Z$ are given by:
\begin{align*}
    Q_{X/Z} = \frac{1}{2} \qty[ 1 - \qty( \frac{ p_i \mp p_x \pm p_z - p_y }{ p_i + p_x + p_z + p_y } )^{L/L_0} ],
\end{align*}
where $p_i$, $p_x$, $p_y$, and $p_z$ are the success, $X$-error, $Y$-error, and $Z$-error probabilities of a CBSM in a single repeater station, respectively.

We also define the effective total cost $Q_{tot}$ of the quantum repeater by:
\begin{align}
    Q_{tot} := \frac{C_{exp}}{Rt_0} \times \frac{L}{L_0}, \label{eq:Q_tot_def}
\end{align}
where $C_{exp}$ is the expected cost of CBSM in a single repeater station defined in \textit{Definition \ref{def:cost_function}}. $Q_{tot}$ quantifies the expected total cost of CBSM to generate a secret key with unit length. In the numerical calculations, we try to find the set of parameters $(n, m, \alpha, j)$ and station interval $L_0$ which minimizes $Q_{tot}$.

\subsubsection{Results}

We find the optimal parameter sets which minimize the effective total cost $Q_{tot}$ for the total distance $L=1000$ km and $L=10000$ km. The parameter sets and the corresponding effective total costs $Q_{tot}$ and expected key lengths $Rt_0$ are:
\begin{align*}
    &L=1000 \text{ km:} \\
    &\qquad (n, m, \alpha, j) = (3, 31, 1.9, 1), ~~ L_0 = 0.7 \text{ km} \\
    &\qquad \rightarrow Q_{tot} = (1.019 \pm 0.003) \times 10^5, \\
    &\qquad \quad~~ Rt_0 = 0.71 \pm 0.02 \\
    &L=10000 \text{ km:} \\
    &\qquad (n, m, \alpha, j) = (5, 41, 1.8, 3), ~~ L_0 = 0.9 \text{ km} \\
    &\qquad \rightarrow Q_{tot} = (2.09 \pm 0.05) \times 10^6, \\
    &\qquad \quad~~ Rt_0 = 0.78 \pm 0.02 \\
\end{align*}

Figure~\ref{fig:eff_total_cost_and_Rt0_against_n_m} shows $Q_{tot}$ and $Rt_0$ of the quantum repeater against the repetition sizes $n$ and $m$ when $L=1000$ km, for different settings of the coherent-state amplitude $\alpha$. Here, $\alpha$, $L_0$, and the letter solidity parameter $j$ are selected to minimize $Q_{tot}$ or maximize $Rt_0$ if they are not fixed explicitly. Figure~\ref{fig:eff_total_cost_and_Rt0_against_n_m}(c) indicates that $Rt_0$ arbitrarily close to unity can be obtained for sufficiently large values of $n$ and $m$. Particularly, $m$ should be sufficiently large to fix $Z$-errors. However, since $X$-errors are very rare compared to failures and $Z$-errors, $n$ does not need to be very large, although it should be larger than 1 to suppress failures.

Comparing the second and third columns of Fig.~\ref{fig:eff_total_cost_and_Rt0_against_n_m}, CBSM with a small value of $\alpha$ requires a relatively large value of $n$ to reach low $Q_{tot}$ and high $Rt_0$. This is the consequence of the fact that BSM of coherent-state qubits with a small value of $\alpha$ has a higher failure probability than that with a large value of $\alpha$, and the effect of failures can be mitigated by increasing $n$ as discussed in Sec.~\ref{subsec:fault_tolerance_of_cbsm}. Meanwhile, the minimal attainable $Q_{tot}$ is smaller for $\alpha=1.9$ than $\alpha=1.4$. The dependence of the performance of the repeater network to $\alpha$ is more clearly shown in Fig.~\ref{fig:eff_total_cost_Rt0_against_alpha_L0}(a). Here, $Q_{tot}$ is minimal at $\alpha=1.9$; this indicates that the parity code with $\alpha > 2.0$ which is hard to generate is unnecessary to attain an efficient repeater.

We also plot the dependence of the optimal $Q_{tot}$ and the corresponding $Rt_0$ to the station interval $L_0$ in Fig.~\ref{fig:eff_total_cost_Rt0_against_alpha_L0}(b). It shows that $Q_{tot}$ is minimal when $L_0$ is around 0.6--1.0 km.

Our repeater scheme shows the similar scale of performance with CBSM based on multi-photon polarization qubits, where $Q_{tot}=6.5 \times 10^4$ and the corresponding key generation rate is 0.70 with the same condition of the total distance and photon loss rate\footnote{For CBSM with multi-photon polarization qubits, we use $C_{exp} = nm$, the number of physical-level BSMs for one CBSM, in the definition of $Q_{tot}$ [Eq.~\eqref{eq:Q_tot_def}].} \cite{lee2019fundamental}, although the precise comparison is impossible due to the difference of the physical-level BSM schemes. Although we cannot say our repeater scheme is better than that in Ref. \cite{lee2019fundamental}, it is still a remarkable result considering that the scheme in Ref. \cite{lee2019fundamental} outperforms recent advanced matter-based and all-optical based schemes \cite{lee2019fundamental}.

\section{Implementation of the modified parity encoding}
\label{sec:implementation}

In this section, we discuss  implementations of the modified parity encoding and its elementary operations. Here, a \textit{logical} gate or measurement means a gate or measurement in modified parity encoding basis $\qty{ \ket{0_L}, \ket{1_L} }$, whereas a \textit{physical} gate or measurement means a gate or measurement in coherent-state basis $\qty{ \ket{\pm\alpha} }$.

\subsection{Logical-level implementations}
\label{subsec:logical_operations}

Here, we investigate the ways to encode a logical qubit and to implement logic gates and measurements, in terms of physical operations.

\begin{figure}[tb]
    \centering
    \begin{quantikz}
        \lstick{$\ket{\psi}$}                   & \ctrl{3} & \ctrl{6} & \targ{}   & \targ{}   & \qw\rstick[wires=9]{$\ket{\psi}_L$} \\
        \lstick{$\ket{\alpha} + \ket{-\alpha}$} & \qw      & \qw      & \ctrl{-1} & \qw       & \qw \\
        \lstick{$\ket{\alpha} + \ket{-\alpha}$} & \qw      & \qw      & \qw       & \ctrl{-2} & \qw \\
        \lstick{$\ket{\alpha}$}                 & \targ{}  & \qw      & \targ{}   & \targ{}   & \qw \\
        \lstick{$\ket{\alpha} + \ket{-\alpha}$} & \qw      & \qw      & \ctrl{-1} & \qw       & \qw \\
        \lstick{$\ket{\alpha} + \ket{-\alpha}$} & \qw      & \qw      & \qw       & \ctrl{-2} & \qw \\
        \lstick{$\ket{\alpha}$}                 & \qw      & \targ{}  & \targ{}   & \targ{}   & \qw \\
        \lstick{$\ket{\alpha} + \ket{-\alpha}$} & \qw      & \qw      & \ctrl{-1} & \qw       & \qw \\
        \lstick{$\ket{\alpha} + \ket{-\alpha}$} & \qw      & \qw      & \qw       & \ctrl{-2} & \qw
    \end{quantikz}
    \caption{Encoding circuit of the modified parity code defined in \textit{Definition \ref{def:parity_encoding}}, for $n=m=3$ case. Here, $\ket{\psi}$ is the desired qubit encoded in coherent-state basis ($\ket{0}=\ket{\alpha}, \ket{1}=\ket{-\alpha}$). and all the controlled-not gates are also under coherent-state basis. $\ket{\alpha} + \ket{-\alpha}$ is the SCS, where the normalization constant is omitted.}
    \label{fig:encoding_circuit}
\end{figure}
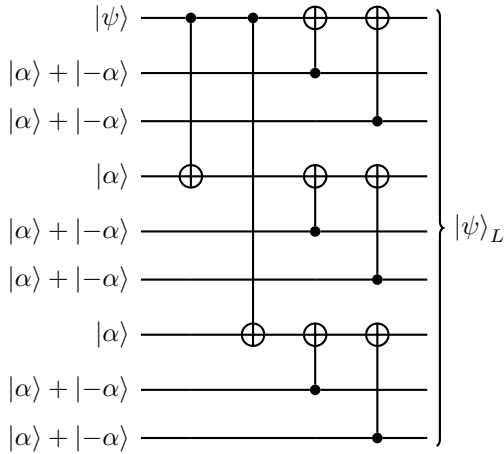

\paragraph*{Encoding.} The encoding circuit of a logical qubit is illustrated in Fig.~\ref{fig:encoding_circuit} for $n=m=3$ case. The desired qubit encoded in coherent-state basis is prepared at the first PLS of the first block. First, controlled-not (CNOT) gates are operated between the first PLS of the first block (control) and the PLSs of the other blocks (target). After that, for each block, CNOT gates are operated between the first PLS (target) and the other ones (control). The encoding circuit for arbitrary values of $n$ and $m$ generally requires $n-1$ copies of the coherent state $\ket{\alpha}$, $n(m-1)$ copies of the SCS $N_+ \qty( \ket{\alpha} + \ket{-\alpha} )$, and $nm - 1$ physical CNOT gates. The decoding circuit is exactly the reverse of the encoding circuit.

\paragraph*{$X_L$ and $Z_L$ gate.} A logical $X$ gate ($X_L$) can be decomposed into $n$ physical $X$ gates, while a logical $Z$ gate ($Z_L$) can be decomposed into $m$ physical $Z$ gates:
\begin{subequations}
\label{eqs:logical_gates_decomposition}
\begin{align}
    X_L &= \prod_{i=1}^n X_{ik} \qquad\text{for any } k \leq n, \label{eq:logical_X_decomposition} \\
    Z_L &= \prod_{k=1}^m Z_{ik} \qquad\text{for any } i \leq m, \label{eq:logical_Z_decomposition}
\end{align}
\end{subequations}
where $X_{ik}$($Z_{ik}$) is a physical $X$($Z$) gate on the $k$th PLS of the $i$th block. $X_L$ and $Z_L$ gates are used in the quantum repeater scheme discussed in Sec.~\ref{subsec:quantum_repeater} to recover the original quantum information from the transmitted state and the classical information on the CBSM results at the end of the network. We note that they are not necessary for the CBSM scheme itself.

\paragraph*{$X_L$ and $Z_L$ measurement.} A $X_L$($Z_L$) measurement is done by the combination of $n$($m$) physical $X$($Z$) measurements as seen in Eqs.~\eqref{eqs:logical_gates_decomposition}. However, this procedure is not fault-tolerant, since a single physical-level $Z$($X$)-error before the measurement or a single physical measurement error causes an error in the measurement. In order to obtain fault-tolerance, one needs to perform multiple measurements for different $k$($i$)'s in Eqs.~\eqref{eqs:logical_gates_decomposition}.


\subsection{Physical-level implementations}

Now, we review the recent progress on implementations of physical-level ingredients required for our scheme including the logical operations discussed in the previous subsection. We need to deal with SCSs, CNOT gates, $X$($Z$) gates, $X$($Z$) measurements, and PNPDs.

\paragraph*{Superpositions of coherent states.} SCSs (often called Schr\"odinger's cat states) in free-propagating optical fields are required for encoding logical qubits. 
It was known that SCSs may be produced using a strong nonlinearity \cite{yurke1986generating} or a precise photon-resolving detector \cite{dakna1997generating, song1990generation} although it was experimentally highly demanding.
Later, the possibilities of generating SCSs using realistic detectors  \cite{lund2004conditional, jeong2005production} or a weak nonlinearity \cite{weak2005, jeong2004generation, jeong2005using} have been explored.
Free-propagating SCSs with amplitudes of $|\alpha| \lessapprox 2$ are now within reach of current technology \cite{ourjoumtsev2007generation,  gerrits2010generation, sychev2017enlargement, asavanant2017generation, serikawa2018generation}. The amplitudes of the generated SCSs are sufficient for our CBSM scheme as discussed in Sec.~\ref{subsec:quantum_repeater}, while their purities are yet to be improved.

\paragraph*{CNOT gates.} Physical CNOT gates are also required for encoding logical qubits. Gate teleportation has been mainly studied for implementing CNOT gates \cite{jeong2002efficient, ralph2003quantum}, which requires particular two-mode or three-mode entangled states as resources and BSMs under coherent-state basis. An alternative way proposed by Marek and Fiurášek employs single-photon subtractions as the driving force, but it has a limitation of being non-deterministic \cite{marek2010elementary}.

\paragraph*{$X$ and $Z$ gates.} Physical $X$($Z$) gates are required for logical $X$($Z$) gates. Implementation of an $X$ gate is straightforward; $\hat{X}=\exp\qty( i\pi\hat{a}^\dagger\hat{a} )$, where $\hat{a}$ is the annihilation operator, is just swifting the electromagnetic wave's phase by $\pi$ \cite{ralph2003quantum}. Implementation of a $Z$ gate is more complicated due to its non-unitarity. An approximate $Z$ gate can be achieved via nonlinear medium \cite{jeong2002efficient}, gate teleportation with resources of SCSs \cite{ralph2003quantum, jeong2007schrodinger}, or single-photon subtraction \cite{marek2010elementary}. The single-photon subtraction method was experimentally demonstrated in \cite{blandino2012characterization}.

\paragraph*{$X$ and $Z$ measurements.} Physical $X$($Z$) measurements are required for logical $X$($Z$) measurements. An $X$ measurement can be approximately implemented via heterodyne measurement \cite{weedbrook2012gaussian}, while a perfect measurement is impossible due to the non-orthogonality between $\ket{\alpha}$ and $\ket{-\alpha}$. A $Z$ measurement is the same as measuring the parity of the photon number, which is what exactly a PNPD does.

\paragraph*{Photon-number parity detectors.} PNPDs are required for physical-level BSMs and $Z$ measurements. There exist two ways to realize a PNPD: detecting the parity of the photon number directly (direct measurement), or detecting it indirectly by measuring the photon number (indirect measurement). Regarding the direct measurement, parity measurements in cavities have been demonstrated and realized via Rydberg atom interacting with photons \cite{haroche2007measuring}, Ramsey interferometry \cite{sun2014tracking, ofek2016extending}, or strong nonlinear Hamiltonian of a Josephson circuit \cite{cohen2017degeneracy}. However, parity measurements of propagating waves have not been covered much yet except a few studies such as parity measurement via strong nonlinear optical switching devices \cite{gerry2005quantum, gerry2010parity} or a cavity QED system realized in superconducting circuits \cite{besse2020parity}. Indirect measurement, or photon-number-resolving (PNR) detection, is a more actively studied topic due to its wide availability \cite{jonsson2019evaluating}. PNR schemes can be classified into two categories: inherent PNR detectors and multiplexed single-photon detectors. Transition edge sensors (TESs) are promising candidates for inherent PNR detectors \cite{cabrera1998detection, miller2003demonstration, lita2008counting, marsili2013detecting, harder2016single}, which can distinguish up to 12 photons with an estimated detection efficiency of 0.98 \cite{sperling2017detector}. While inherent PNR detectors generally demand tricky conditions \cite{lubin2019quantum}, multiplexed single-photon detectors exploit several inexpensive single-photon detectors \cite{fitch2003photon, achilles2003fiber, achilles2004photon, divochiy2008superconducting, mattioli2015photon, nehra2020photon}. However, it is currently difficult to achieve a sufficiently high efficiency with multiplexed single-photon detectors, e.g., one cannot resolve more than three photons with better-than-guessing quality using ideal click detectors with an eight-segment detector \cite{jonsson2019evaluating}.


\section{Conclusion}
\label{sec:conclusion}

Bell-state measurement (BSM) is an essential element for optical quantum information processing, particularly for long-range communication through a quantum repeater. The original coherent-state qubit with basis $\qty{ \ket{\pm\alpha} }$ enables one to perform nearly deterministic BSM, but it is vulnerable to dephasing by photon loss especially for large values of amplitude $\alpha$ of coherent states required to reduce non-orthogonality. Fault-tolerant operations with encoded coherent-state qubits have been studied mainly with cavity systems, but this cannot be directly applied to free-propagating fields. 

In this paper, we have explored the possibility to use such encoded coherent-state qubits for long-range quantum communication by designing an appropriate encoding scheme and fault-tolerant BSM scheme. We have presented the modified parity encoding which is a natural extension of the original coherent-state encoding, and also suggested a hardware-efficient concatenated Bell-state measurement (CBSM) scheme in a completely or partially distributed manner. We have argued and numerically verified that the CBSM scheme successfully suppresses both failures and dephasing simultaneously. We have also shown that SCSs with reasonable values of the amplitude such as $\alpha \lessapprox 2$ are enough to achieve the success probability close to unity. It is worth noting this point since it is difficult to generate superpositions of coherent states (SCSs) with large amplitudes. It is known that free-propagating SCSs with $\alpha \lessapprox 2$ can be generated using current technology \cite{ourjoumtsev2007generation,  gerrits2010generation, sychev2017enlargement}. Furthermore, we have shown that the quantum repeater scheme using the CBSM scheme for error correction enables efficient long-range quantum communication over 1000 km, where the performance against the cost is on a similar level with the CBSM scheme of multi-photon polarization qubit \cite{lee2019fundamental}.

In summary, we have demonstrated that the properly encoded coherent-state qubits in free-propagating fields provide an alternative way for fault-tolerant information processing enabling long-range communication. In addition to presenting the possibility, we have shown that the performance of our CBSM and repeater scheme is comparable to that of other methods, or even outperforms for some cases.

Our encoding and CBSM schemes are relatively simple. The modified parity encoding is a simple generalized Shor's 9-qubit code and the CBSM is also just a classical information processing with the results of well-known physical-level BSMs. We have further shown that the methods to encode logical qubits and implement logical gates and measurements are elementary compositions of physical-level gates or measurements in the coherent-state basis. Therefore, the most challenging part to realize our scheme is on the physical level, such as generating free-propagating SCSs, elementary logical gates/measurements, and photon-number parity detectors (PNPDs) used for physical-level BSM. Fortunately, a number of appropriate implementation methods have been proposed for all of them, even though some of them are non-deterministic or costly. Furthermore, one remarkable point is that logical gates on coherent-state qubits can be implemented with linear-optical devices and off-line production of resource states \cite{jeong2002efficient, ralph2003quantum}; thus, only linear-optical devices are used during BSM while the required resources are generated beforehand. 
Of course, experimental imperfections during these physical-level processes would also affect the performance of our scheme, and details of such effects require further investigations as future work.

\section*{Acknowledgments}
This work was supported by National Research Foundation of Korea grants funded by the Korea government (NRF-2020R1A2C1008609, NRF-2019M3E4A1080074 and NRF-2020K2A9A1A06102946) via the Institute for Applied Physics at Seoul National University. S.W.L. acknowledges support from the National Research Foundation of Korea (2020M3E4A1079939) and the KIST institutional program (2E31021).

\appendix

\section{Positive-operator valued measure elements of Bell-state measurement on coherent-state qubits in lossy environment}
\label{app:povm_elements}

Here, we explicitly present the positive-operator valued measure (POVM) elements of BSM under the basis of $\qty{ \ket{\pm\alpha} }$ in lossy environment, which is dealt in Sec.~\ref{sec:bsm_coherent_state_qubits}. The set of operators $\qty{ M_{x,y} \middle| x, y \in \{ 0, 1, 2 \} }$ where
\begin{equation*}
    M_{x,y} := \qty[ \mathcal{U}_{\mathrm{BS}} \circ \qty( \mathit{\Lambda}_{\eta_1} \otimes \mathit{\Lambda}_{\eta_2} ) ]^\dagger \qty( \Pi_x \otimes \Pi_y )
\end{equation*}
forms a POVM corresponding to the BSM of coherent-state qubits, where $\mathcal{U}_{\mathrm{BS}}$ is a unitary channel corresponding to a 50:50 beam splitter, $\mathit{\Lambda}_{\eta}$ is a photon loss channel with a survival rate of $\eta$, and $\Pi_x$ is a projector defined by
\begin{align*}
    &\Pi_0 := \dyad{0_\mathrm{F}}, \quad \Pi_1 := \sum_{n \geq 1:\mathrm{odd}} \dyad{n_\mathrm{F}},\\
    &\Pi_2 := \sum_{n \geq 2:\mathrm{even}} \dyad{n_\mathrm{F}},
\end{align*}
where $\ket{n_\mathrm{F}}$ is the Fock state with a photon number of $n$. The photon loss channel $\Lambda_\eta$ transforms $\dyad{\alpha}$ and $\dyad{\alpha}{-\alpha}$ as follows:
\begin{align}
    \Lambda_\eta \qty( \dyad{\alpha} ) &= \dyad{\sqrt{\eta}\alpha}, \nonumber \\
    \Lambda_\eta \qty( \dyad{\alpha}{-\alpha} ) &= e^{-2(1-\eta)|\alpha|^2} \dyad{\sqrt{\eta}\alpha}{-\sqrt{\eta}\alpha}. \label{eq:transformation_of_cross_term_by_photon_loss}
\end{align}
With these relations, we find the analytic expressions of the matrix elements of each POVM element $M_{x,y}$ as:
\begin{subequations}
\label{eqs:M_xy}
\begin{align}
    &\mel{\phi_\pm}{M_{x,y}}{\phi_\pm} = c_\pm \qty[ 1 \pm (-1)^{x+y} e^{-2\qty( 2 - \eta_1 - \eta_2 ) |\alpha|^2} ] \nonumber \\
    &\qquad\qquad\qquad\quad~~ \times f_x \qty( \eta_+ ) f_y \qty( \eta_-), \label{eq:M_xy_phi_phi} \\
    &\mel{\psi_\pm}{M_{x,y}}{\psi_\pm} = c_\pm \qty[ 1 \pm (-1)^{x+y} e^{-2\qty( 2 - \eta_1 - \eta_2 ) |\alpha|^2} ] \nonumber \\
    &\qquad\qquad\qquad\quad~~ \times f_x \qty( \eta_- ) f_y \qty( \eta_+  ), \label{eq:M_xy_psi_psi} \\
    &\mel{\phi_\pm}{M_{x,y}}{\psi_\pm} \nonumber \\
    &\quad = c_\pm \qty[ \pm (-1)^{x+y} e^{ -2 \qty( 1 - \eta_1 ) |\alpha|^2 } + e^{ -2 \qty( 1 - \eta_2 ) |\alpha|^2 } ] \nonumber \\
    &\quad\quad \times f_x \qty( \sqrt{\eta_+ \eta_-} ) f_y \qty( \sqrt{\eta_+ \eta_-} ), \label{eq:M_xy_phi_psi} \\
    &\mel{\phi_+}{M_{x,y}}{\psi_-} = \mel{\psi_+}{M_{x,y}}{\psi_-} = \mel{\phi_\pm}{M_{x,y}}{\psi_\mp} \nonumber \\
    &\quad = 0, \label{eq:M_xy_vanished}
\end{align}
\end{subequations}
where
\begin{align*}
    c_\pm := \frac{e^{-(\eta_1 + \eta_2)|\alpha|^2}}{1 \pm e^{-4|\alpha|^2}}, \quad \eta_\pm := \frac{\qty( \sqrt{\eta_1} \pm \sqrt{\eta_2} )^2}{2}, \\
    f_i (\eta) := 
        \begin{cases}
            1 & \text{if } i = 0, \\
            \sinh \qty( \eta |\alpha|^2 ) & \text{if } i = 1, \\
            \cosh \qty( \eta |\alpha|^2 ) - 1 & \text{if } i = 2.
        \end{cases}
\end{align*}

\section{Derivation of the probability distributions of concatenated Bell-state measurement results}
\label{app:prob_dist_of_measurement_results}

In this appendix, we show a brief outline to induce the analytic expressions of the probability distributions of CBSM results conditioning to the initial Bell states before the measurement. We only consider the unoptimized CBSM scheme, since the measurement results of the hardware-efficient CBSM scheme are the direct consequences of those of the unoptimized scheme.

\subsection{Derivation of the probability distributions of block-level results}
\label{subapp:prob_dist_of_bsm_1_results}

We first find the probability distributions of block-level BSM results, conditioning to the initial block-level Bell state. A single \bsm{1} result can be expressed by two vectors $\vb{x}, \vb{y} \in \{0, 1, 2, 3 \}^m$, where the $i$th elements of them are the two PNPD results of the $i$th PLS. We want to find $\P(\vb{x}, \vb{y} | B_1)$ for $\ket{B_1} \in \mathcal{B}_1 := \qty{ \ket{\phi^{(m)}_\pm}, \ket{\psi^{(m)}_\pm} }$

From Eqs.~\eqref{eq:single_bsm_cond_prob} and \eqref{eq:block_to_physical_phi}, the conditional probability for the initial state of $\ket{B_1} = \ket{\phi^{(m)}_\pm}$ is express as:
\begin{align}
    &\P(\vb{x}, \vb{y} | \phi^{(m)}_\pm) = \bra{\phi^{(m)}_\pm} \bigotimes_{i=1}^m M_{x_i, y_i} \ket{\phi^{(m)}_\pm} \nonumber \\
    &\quad = \frac{1}{2} \tilde{N}_{\pm}(1, m)^2 \sum_{l,l'=\text{even} \leq m} g^\pm_{m,l,l'} (\vb{x}, \vb{y}), \label{eq:block_cond_prob_expressed_by_g}
\end{align}
where $\tilde{N}_\pm (1, m)$ is defined in Eq.~\eqref{eq:N_tilde_def}. The function $g^\pm_{m,l,l'} (\vb{x}, \vb{y})$ is defined as:
\begin{align*}
    &g^\pm_{m,l,l'} (\vb{x}, \vb{y}) := \sum_{\substack{\bigotimes_{i=1}^m \ket{P_i} \in\text{Perm}\qty[ \ket{\psi_\pm}^{\otimes l} \ket{\phi_\pm}^{\otimes m-l} ] \\ \bigotimes_{i=1}^m \ket{P'_i} \in \text{Perm}\qty[ \ket{\psi_\pm}^{\otimes l'} \ket{\phi_\pm}^{\otimes m-l'} ]}} \\
    &\qquad\qquad\qquad\qquad\qquad\qquad \qty[ \prod_{i=1}^m \bra{P_i} M_{x_i, y_i} \ket{P'_i} ],
\end{align*}
where $\text{Perm}[\cdot]$ is the set of all the permutations of tensor products inside the square bracket. The function $g^\pm_{m,l,l'}$ has a recurrence relation: (omit $\vb{x}$ and $\vb{y}$ for simplicity)
\begin{align}
    g^\pm_{m,l,l'} =~& g^\pm_{m-1,l,l'} M^{(m)\pm}_{11} \nonumber \\
    &+ \qty[ g^\pm_{m-1,l,l'-1} + g^\pm_{m-1,l-1,l'} ] M^{(m)\pm}_{12} \nonumber \\
    &+ g^\pm_{m-1,l-1,l'-1} M^{(m)\pm}_{22} \label{eq:block_recur_g}
\end{align}
where
\begin{subequations}
\label{eqs:M_abbrv_def}
\begin{align}
    M^{(k)\pm}_{11} &:= \mel{\phi_\pm}{\hat{M}_{x_k, y_k}}{\phi_\pm}, \label{eq:M_11_abbrv_def} \\
    M^{(k)\pm}_{12} &:= \mel{\phi_\pm}{\hat{M}_{x_k, y_k}}{\psi_\pm}, \label{eq:M_12_abbrv_def} \\
    M^{(k)\pm}_{22} &:= \mel{\psi_\pm}{\hat{M}_{x_k, y_k}}{\psi_\pm}, \label{eq:M_22_abbrv_def}
\end{align}
\end{subequations}
which can be calculated from Eqs.~\eqref{eqs:M_xy}. Now, we define a vector $\vb{v}^\pm_m (\vb{x}, \vb{y})$: (Note that $g_{m,l,l'}^\pm$ is a function of $\vb{x}$ and $\vb{y}$.)
\begin{align}
  \vb{v}^\pm_m := &\left( \sum_{l,l':\text{even} \leq m} g^\pm_{m,l,l'},~ \sum_{\substack{l:\text{even} \leq m \\ l':\text{odd} \leq m}} g^\pm_{m,l,l'}, \right. \nonumber \\
  &\quad \left. \sum_{\substack{l:\text{odd} \leq m \\ l':\text{even} \leq m}} g^\pm_{m,l,l'},~ \sum_{l,l':\text{odd} \leq m} g^\pm_{m,l,l'} \right)^T \label{eq:vector_v_def}
\end{align}
From Eq.~\eqref{eq:block_recur_g}, we get a recurrence relation of $\vb{v}^\pm_m$:
\begin{align}
    \vb{v}^\pm_m &=
        \begin{pmatrix}
            M^{(m)\pm}_{11} & M^{(m)\pm}_{12} & M^{(m)\pm}_{12} & M^{(m)\pm}_{22} \\
            M^{(m)\pm}_{12} & M^{(m)\pm}_{11} & M^{(m)\pm}_{22} & M^{(m)\pm}_{12} \\
            M^{(m)\pm}_{12} & M^{(m)\pm}_{22} & M^{(m)\pm}_{11} & M^{(m)\pm}_{12} \\
            M^{(m)\pm}_{22} & M^{(m)\pm}_{12} & M^{(m)\pm}_{12} & M^{(m)\pm}_{11}
        \end{pmatrix}
    \vb{v}^\pm_{m-1} \nonumber \\
    &:= \vb{\tilde{M}}^\pm_{x_m,y_m} \vb{v}^\pm_{m-1}. \label{eq:v_recurrence}
\end{align}
Considering the initial condition at $m=1$, $\vb{v}^\pm_m (\vb{x}, \vb{y})$ is written as:
\begin{align}
    \vb{v}^\pm_m (\vb{x}, \vb{y}) = \vb{\tilde{M}}^\pm_{x_m,y_m} \cdots \vb{\tilde{M}}^\pm_{x_1,y_1} (1,~0,~0,~0)^T. \label{eq:block_v_expression}
\end{align}
Finally, $\P(\vb{x}, \vb{y} | \phi^{(m)}_\pm)$ is written in terms of the vector $\vb{v}^\pm_m$ using Eqs.~\eqref{eq:block_cond_prob_expressed_by_g} and \eqref{eq:vector_v_def}:
\begin{subequations}
\label{eqs:block_cond_probs_by_v}
\begin{align}
    \P(\vb{x}, \vb{y}|\phi^{(m)}_\pm) &= \frac{1}{2} \tilde{N}_{\pm}(1, m)^2  v^\pm_{m1} (\vb{x}, \vb{y}). \label{eq:block_cond_prob_phi_by_v},
\end{align}
where $v_{mi}^\pm$ is the $i$th element of $\vb{v}_m^\pm$. In the similar way, $\P(\vb{x}, \vb{y} | \psi^{(m)}_\pm)$ is written as:
\begin{align}
    \P(\vb{x}, \vb{y}|\psi^{(m)}_\pm) &= \frac{1}{2} \tilde{N}_{\pm}(1, m)^2 v^\pm_{m4} (\vb{x}, \vb{y}). \label{eq:block_cond_prob_psi_by_v}
\end{align}
\end{subequations}

In conclusion, the conditional probability distribution of CBSM results conditioning to the input block-level Bell state is obtained from Eqs.~\eqref{eqs:block_cond_probs_by_v} with Eqs.~\eqref{eqs:M_xy}, \eqref{eqs:M_abbrv_def}, \eqref{eq:v_recurrence}, and \eqref{eq:block_v_expression}, all of which are written in simple matrix forms.

\subsection{Derivation of the probability distributions of logical-level results}
\label{subapp:prob_dist_of_bsm_2_results}

Now, we consider the probability distributions of logical-level results conditioning to the initial logical-level Bell state, which is the goal of this appendix. A single CBSM result can be expressed by two matrices $\vb{X}, \vb{Y} \in \{ 0, 1, 2, 3 \}^{n \times m}$, where the $(i,k)$ elements of them are the two PNPD results of the $k$th PLS of the $i$th block. What we want to find is the probability distribution $\P(\vb{X}, \vb{Y} | B_2)$ for $\ket{B_2} \in \mathcal{B}_2 := \qty{ \ket{\Phi_\pm}, \ket{\Psi_\pm} }$.

Because of the similarity of Eqs.~\eqref{eqs:logical_to_block} and \eqref{eqs:block_to_physical}, we can follow the almost same logical structure with the previous subsection when finding the expressions of the probability distributions. However, there exist three main differences between the block and logical level. First, the roles of the letters and signs are inverted between the two sets of the equations. Second, there are unnormalized states in the summations of Eqs.~\eqref{eqs:logical_to_block}, unlike Eqs.~\eqref{eqs:block_to_physical}. Lastly, $\mel{ \phi_+^{(m)} \qty(\psi_+^{(m)}) }{  \bigotimes_{k=1}^m \hat{M}_{x_k, y_k} }{ \phi_-^{(m)} \qty(\psi_-^{(m)}) }$ vanish unlike the corresponding one in block level, i.e., $\mel{\phi_\pm}{\hat{M}_{x, y}}{\psi_\pm}$ in Eq.~\ref{eq:M_12_abbrv_def}.

Considering the differences, we define $2 \times 2$ matrices $\vb{\tilde{L}}^{\phi}_{\vb{x},\vb{y}}$ and $\vb{\tilde{L}}^{\psi}_{\vb{x},\vb{y}}$ where $\vb{x}, \vb{y} \in \qty{ 0, 1, 2, 3 }^m$, instead of $4 \times 4$ matrices, in the similar way with the block-level case:
\begin{align*}
    \vb{\tilde{L}}^{\phi(\psi)}_{\vb{x}, \vb{y}} :=
        \begin{pmatrix}
            L^{\phi(\psi)}_{+} & L^{\phi(\psi)}_{-} \\
            L^{\phi(\psi)}_{-} & L^{\phi(\psi)}_{+}
        \end{pmatrix},
\end{align*}
where
\begin{align}
\label{eq:L_def_result_app}
    &L^{\phi(\psi)}_\pm := \qty[ 1 \pm u(\alpha, m)^2 ] \nonumber \\
    &\quad \times \expval{ \bigotimes_{k=1}^m \hat{M}_{x_k, y_k} }{ \phi^{(m)}_\pm \qty( \psi^{(m)}_\pm ) },
\end{align}
$u(\alpha, m)$ is defined in Eq.~\eqref{eq:u_def}, and $x_k$($y_k$) is the $k$th element of $\vb{x}$($\vb{y}$). 
We do not need $4 \times 4$ matrices since the off-diagonal elements of $\bigotimes_{k=1}^m \hat{M}_{x_k, y_k}$ between two Bell states of different signs vanish. 
We also note that the RHS of Eq.~\eqref{eq:L_def_result_app} can be calculated from Eqs.~\eqref{eqs:block_cond_probs_by_v}. 
The conditional probability $\P(\vb{X}, \vb{Y} | B_2)$, where the $i$th row vector of $\vb{X}$($\vb{Y}$) is $\vb{x}_i$($\vb{y}_i$), is then
\begin{subequations}
\label{eqs:log_cond_probs_by_w}
\begin{align}
    \P(\vb{X}, \vb{Y}|\Phi_+ (\Psi_+)) &= \tilde{N}_{+}(n, m)^2 w^{\phi (\psi)}_{n1} (\vb{X}, \vb{Y}), \\
    \P(\vb{X}, \vb{Y}|\Phi_- (\Psi_-)) &= \tilde{N}_{-}(n, m)^2 w^{\phi (\psi)}_{n2} (\vb{X}, \vb{Y}),
\end{align}
\end{subequations}
where $\tilde{N}_\pm (n, m)$ is defined in Eq.~\eqref{eq:N_tilde_def} and $w_{n\mu}^{\phi(\psi)} (\vb{X}, \vb{Y})$ is the $\mu$th element of the two-dimensional vector $\vb{w}^{\phi(\psi)}_n (\vb{X}, \vb{Y})$ defined by:
\begin{align}
\label{eq:logical_w_definition}
    \vb{w}^{\phi(\psi)}_n (\vb{X}, \vb{Y}) := \vb{\tilde{L}}^{\phi(\psi)}_{\vb{x}_n, \vb{y}_n} \cdots \vb{\tilde{L}}^{\phi(\psi)}_{\vb{x}_1, \vb{y}_1} (1, 0)^T.
\end{align}

\section{Method for sampling concatenated Bell-state measurement results}
\label{app:method_sampling_measurement_results}

In this appendix, we explain the method to sample CBSM results. Since we have the analytic expressions of the probability distributions of measurement results [Eqs.~\eqref{eqs:log_cond_probs_by_w}], it is possible to sample arbitrary CBSM results, each of which is composed of $2nm$ PNPD results. However, since the number of CBSM results increases exponentially on $n$ and $m$, it is computationally expensive to use this method. Instead of that, denoting $(p,q)$ the $q$th PLS of the $p$th block, we sample the results for each PLS in order: $(1,1) \rightarrow (1,2) \rightarrow (1,3) \rightarrow \cdots \rightarrow (1,m) \rightarrow (2,1) \rightarrow \cdots \rightarrow (2,m) \rightarrow \cdots \rightarrow (n,m)$. Therefore, we need the conditional probability of getting each $(p,q)$ result conditioning to all the results before $(p,q)$.

The conditional probability we want is
\begin{align}
    &\P(x_{pq}, y_{pq}|x_{11}, y_{11}, \cdots, x_{p'q'}, y_{p'q'}; B_2) \nonumber \\
    &\quad \propto \P(x_{11}, y_{11}, \cdots, x_{pq}, y_{pq} | B_2) \nonumber \\
    &\quad = \mathbf{Pr} \left( \vb{x}_1, \vb{y}_1, \cdots, \vb{x}_{p-1}, \vb{y}_{p-1}, \right. \nonumber \\
    &\qquad\qquad \left. x_{p1}, y_{p1}, \cdots, x_{pq}, y_{pq} \middle| B_2 \right), \label{eq:cond_prob_to_joint_prob}
\end{align}
where
\begin{align*}
    (p',q') =
        \begin{cases}
            (p, q-1) &\text{if } q > 1, \\
            (p-1, m) &\text{otherwise},
        \end{cases}
\end{align*}
$x_{ik}$ and $y_{ik}$ are the measurement results of the two $(i,k)$ PNPDs, $\ket{B_2} \in \mathcal{B}_2$, and $\vb{x}_i$ ($\vb{y}_i$) is a vector whose $k$th element is $x_{ik}$ ($y_{ik}$). Note that the proportionality is valid only when $x_{11}$, $y_{11}$, ..., $x_{p'q'}$, $y_{p'q'}$ are fixed. From now on, we use the proportionality notation while assuming this condition. Using Eq.~\eqref{eq:logical_to_block_phi} and the fact from Eq.~\eqref{eq:M_xy_vanished} that the cross terms of $\bigotimes_{s=1}^m M_{x_{rs},y_{rs}}$ between Bell states with different signs vanish, it is deduced that the RHS of Eq.~\eqref{eq:cond_prob_to_joint_prob} with $\ket{B_2} = \ket{\Phi_\pm}$ is
\begin{align}
    &\mathbf{Pr} \left( \vb{x}_1, \vb{y}_1, \cdots, \vb{x}_{p-1}, \vb{y}_{p-1}, \right. \nonumber \\
    &\qquad\left. x_{p1}, y_{p1}, \cdots, x_{pq}, y_{pq} \middle| \Phi_{+(-)} \right) \nonumber \\
    &\quad \propto \sum_{k:\text{even(odd)} \leq n} \sum_{ \bigotimes_{r=1}^n \ket{P^{(m)}_r} \in \text{Perm}\qty[ \ket{\Phi_-}^{\otimes k} \ket{\Phi_+}^{\otimes n-k} ] } \nonumber \\
    &\qquad\qquad\qquad\qquad\quad \qty[ C_+^{2(n-k)}C_-^{2j} h^{(p,q)}_{P_1, \cdots, P_p} ], \label{eq:cond_prob_to_previous_results}
\end{align}
where $C_\pm := \qty[ 1 \pm u(\alpha,m)^2 ]^{1/2}$, the normalization constant in Eq.~\eqref{eq:phi_tilde_def}. Also,
\begin{align}
    &h^{(p,q)}_{P_1, \cdots, P_p} := \qty[ \prod_{r=1}^{p-1} \mel{ P^{(m)}_r }{ \bigotimes_{s=1}^m M_{rs} }{ P^{(m)}_r } ] \nonumber \\
    &\qquad \times \mel{ P^{(m)}_p }{ \qty( \bigotimes_{s=1}^q M_{ps} ) \otimes I^{\otimes m-q} }{ P^{(m)}_p }, \label{eq:f_definition}
\end{align}
where $M_{rs} := M_{x_{rs},y_{rs}}$ and $I$ is the identity operator in a single PLS. Again, using Eq.~\eqref{eq:block_to_physical_phi}, the last part of the RHS of the above definition is:
\begin{align*}
    &\mel{ \phi_\pm^{(m)} }{ \qty( \bigotimes_{s=1}^q M_{ps} ) \otimes I^{\otimes m-q} }{ \phi_\pm^{(m)} } \\
    &\quad \propto \sum_{l,l':\text{even} \leq m} \sum_{\substack{\bigotimes_{s=1}^m \ket{P_s} \in \text{Perm}\qty[ \ket{\psi_\pm}^{\otimes l} \ket{\phi_\pm}^{\otimes m-l} ] \\ \bigotimes_{s=1}^m \ket{P'_s} \in \text{Perm}\qty[ \ket{\psi_\pm}^{\otimes l'} \ket{\phi_\pm}^{\otimes m-l'} ] }} \\
    &\qquad\qquad\qquad\qquad \qty[ \prod_{s=1}^q \mel{P_s}{M_{ps}}{P'_s} \prod_{s=q+1}^m \braket{P_s}{P'_s} ] \\
    &\quad := \xi^\phi_{\pm, p, q}.
\end{align*}
After transforming the RHS of the above equation appropriately with using the fact that $\bra{\phi_-} \ket{\psi_-}$ vanishes while $\bra{\phi_+}\ket{\psi_+}$ does not, we obtain:
\begin{subequations}
\label{eqs:xi_phi_results}
\begin{align}
    \xi^\phi_{+, p, q} &=
        \begin{cases}
            R_q^+ \qty( v^+_{q1} + v^+_{q4} ) \\
            \quad + R_q^- \qty( v^+_{q2} + v^+_{q3} ) & \text{if } q < m, \\
            v^+_{m1} & \text{if } q = m,
        \end{cases} \label{eq:xi_phi_plus_result} \\
    \xi^\phi_{-, p, q} &=
        \begin{cases}
            v^-_{q1} + v^-_{q4} & \text{if } q < m, \\
            v^-_{m1} & \text{if } q = m,
        \end{cases} \label{eq:xi_phi_minus_result}
\end{align}
\end{subequations}
where $R_q^\pm := (1 + \bra{\phi_+} \ket{\psi_+})^{m-q} \pm (1 - \bra{\phi_+} \ket{\psi_+})^{m-q}$ and $v^\pm_{qi}$ is the $i$th element of vector $\vb{v}^\pm_q \qty( x_1, y_1, \cdots, x_p, y_p )$ calculated from Eq.~\eqref{eq:block_v_expression}. Substituting these on Eq.~\eqref{eq:f_definition} and transforming it appropriately, Eqs.~\eqref{eq:cond_prob_to_joint_prob} and \eqref{eq:cond_prob_to_previous_results} become:
\begin{align}
    &\P(x_{pq}, y_{pq}|x_{11}, y_{11}, \cdots, x_{p'q'}, y_{p'q'}; \Phi_\pm) \nonumber \\
    &\quad \propto \mathbf{Pr} \left( \vb{x}_1, \vb{y}_1, \cdots, \vb{x}_{p-1}, \vb{y}_{p-1}, \right. \nonumber \\
    &\qquad\qquad \left. x_{p1}, y_{p1}, \cdots, x_{pq}, y_{pq} \middle| \Phi_{\pm} \right) \nonumber \\
    &\quad \propto
        \begin{cases}
            C_+^2 \xi^\phi_{+,p,q} \qty( D^\pm_p w^\phi_{p-1,1} + D^\mp_p w^\phi_{p-1,2} ) \\
            \quad + C_-^2 \xi^\phi_{-,p,q} \qty( D^\pm_p w^\phi_{p-1,2} + D^\mp_p w^\phi_{p-1,1} ) \\
            \qquad\qquad\qquad\qquad\qquad\qquad\qquad \text{if } p<n, \\
            C_\pm^2 \xi^\phi_{\pm,p,q} w^\phi_{n-1,1} + C_\mp^2 \xi^\phi_{\mp,p,q} w^\phi_{n-1,2} \\
            \qquad\qquad\qquad\qquad\qquad\qquad\qquad \text{if } p=n,
        \end{cases} \label{eq:cond_prob_phi_final_result}
\end{align}
where $D^\pm_p := \qty( C_+^2 + C_-^2 )^{n-p} \pm \qty( C_+^2 - C_-^2 )^{n-p}$ and $w^\phi_{p-1,i}$ is the $i$th element of vector $\vb{w}^\phi_{p-1} \qty( \vb{x}_{1}, \vb{y}_{1}, \cdots, \vb{x}_{p-1}, \vb{y}_{p-1} )$ defined in Eqs.~\eqref{eq:logical_w_definition}.

The probability distribution for the initial state of $\ket{B_2}=\ket{\Psi_\pm}$ is obtained in very similar way with the above arguments. The result is as follows:
\begin{align}
    &\P(x_{pq}, y_{pq}|x_{11}, y_{11}, \cdots, x_{p'q'}, y_{p'q'}; \Psi_\pm) \nonumber \\
    &\quad \propto
        \begin{cases}
            C_+^2 \xi^\psi_{+,p,q} \qty( D^\pm_p w^\psi_{p-1,1} + D^\mp_p w^\psi_{p-1,2} ) \\
            \quad + C_-^2 \xi^\psi_{-,p,q} \qty( D^\pm_p w^\psi_{p-1,2} + D^\mp_p w^\psi_{p-1,1} ) \\
            \qquad\qquad\qquad\qquad\qquad\qquad\qquad \text{if } p<n, \\
            C_\pm^2 \xi^\psi_{\pm,p,q} w^\psi_{n-1,1} + C_\mp^2 \xi^\psi_{\mp,p,q} w^\psi_{n-1,2} \\
            \qquad\qquad\qquad\qquad\qquad\qquad\qquad \text{if } p=n,
        \end{cases} \label{eq:cond_prob_psi_final_result}
\end{align}
where $w^\psi_{p-1,i}$ is the $i$th element of vector $\vb{w}^\psi_{p-1} \qty( \vb{x}_{1}, \vb{y}_{1}, \cdots, \vb{x}_{p-1}, \vb{y}_{p-1} )$ defined in Eq.~\eqref{eq:logical_w_definition}, and
\begin{subequations}
\label{eqs:xi_psi_results}
\begin{align}
    \xi^\psi_{+, p, q} &=
        \begin{cases}
            R_q^+ \qty( v^+_{q1} + v^+_{q4} ) \\
            \quad + R_q^- \qty( v^+_{q2} + v^+_{q3} ) & \text{if } q < m, \\
            v^+_{m4} & \text{if } q = m,
        \end{cases} \label{eq:xi_psi_plus_result} \\
    \xi^\psi_{-, p, q} &=
        \begin{cases}
            v^-_{q1} + v^-_{q4} & \text{if } q < m, \\
            v^-_{m4} & \text{if } q = m.
        \end{cases} \label{eq:xi_psi_minus_result}
\end{align}
\end{subequations}

In summary, the probability distributions of $(p,q)$ results conditioning to the previous measurement results $(1,1), \cdots, (p',q')$ and the initial logical-level Bell state can be obtained from Eqs.~\eqref{eq:cond_prob_phi_final_result} and \eqref{eq:cond_prob_psi_final_result} together with Eqs.~\eqref{eqs:xi_phi_results} and \eqref{eqs:xi_psi_results}. We use these probability distributions to sample each physical level one by one in order. There are only nine possible results for each PLS and the number of PLSs increases linearly on $n$ and $m$. Hence, it is exponentially fast comparing to sampling the results with total joint probabilities.

\bibliography{references}

\end{document}